\documentclass[letterpaper,
twocolumn,
accepted=2025-04-06
]{quantumarticle}
\pdfoutput=1

\usepackage{caption}
\usepackage{graphicx}	
\usepackage{mathtools}
\usepackage{amsfonts}
\usepackage{amssymb}
\usepackage{physics}
\usepackage{slashed}
\usepackage{circuitikz}
\usepackage{import}
\usepackage{subfig}
\usepackage[utf8]{inputenc}
\usepackage[colorlinks=true]{hyperref}
\usepackage[capitalise]{cleveref}
\usepackage{accents}
\usepackage{orcidlink}
\usepackage[normalem]{ulem}
\usepackage[english]{babel}
\usepackage[T1]{fontenc}

\newcommand{\vs}{\vec{s}}
\newcommand{\vsp}{\vec{s}\,^\prime}

\usepackage[square,numbers,comma,sort&compress]{natbib}
\begin{document}


\title{Simulating Meson Scattering on Spin Quantum Simulators }

\author{
  Elizabeth~R.~Bennewitz
}
\email[
Corresponding author: ]{ebennewi@umd.edu}
\affiliation{Joint Center for Quantum Information and Computer Science, NIST/University of Maryland, College Park, MD 20742 USA}
\affiliation{Joint Quantum Institute, NIST/University of Maryland, College Park, MD 20742 USA}

\author{Brayden~Ware}
\affiliation{Joint Center for Quantum Information and Computer Science, NIST/University of Maryland, College Park, MD 20742 USA}
\affiliation{Joint Quantum Institute, NIST/University of Maryland, College Park, MD 20742 USA}

\author{Alexander~Schuckert}
\affiliation{Joint Center for Quantum Information and Computer Science, NIST/University of Maryland, College Park, MD 20742 USA}
\affiliation{Joint Quantum Institute, NIST/University of Maryland, College Park, MD 20742 USA}

\author{Alessio~Lerose}
\affiliation{Department of Theoretical Physics,
University of Geneva, Quai Ernest-Ansermet 30,
1205 Geneva, Switzerland}
\affiliation{Rudolf Peierls Centre for Theoretical Physics, Clarendon Laboratory, Oxford OX1 3PU, United Kingdom}

\author{Federica~M.~Surace}
\affiliation{Department of Physics and Institute for Quantum Information and Matter, California Institute of Technology, Pasadena, California 91125, USA}

\author{Ron~Belyansky}
\affiliation{Joint Center for Quantum Information and Computer Science, NIST/University of Maryland, College Park, MD 20742 USA}
\affiliation{Joint Quantum Institute, NIST/University of Maryland, College Park, MD 20742 USA}
\affiliation{Pritzker School of Molecular Engineering, The University of Chicago, Chicago, Illinois 60637, USA}

\author{William~Morong}
\thanks{Current address: AWS Center for Quantum Computing, Pasadena, California 91125, USA. Work done prior to joining AWS.}
\affiliation{Joint Center for Quantum Information and Computer Science, NIST/University of Maryland, College Park, MD 20742 USA}
\affiliation{Joint Quantum Institute, NIST/University of Maryland, College Park, MD 20742 USA}

\author{De~Luo}
\affiliation{Duke Quantum Center, Department of Physics and Electrical and Computer Engineering, Duke University, Durham, NC 27701 USA}

\author{Arinjoy~De\orcidlink{0000-0001-9184-8434}} 
\affiliation{Joint Center for Quantum Information and Computer Science, NIST/University of Maryland, College Park, MD 20742 USA}
\affiliation{Joint Quantum Institute, NIST/University of Maryland, College Park, MD 20742 USA}

\author{Kate~S.~Collins}
\affiliation{Joint Center for Quantum Information and Computer Science, NIST/University of Maryland, College Park, MD 20742 USA}
\affiliation{Joint Quantum Institute, NIST/University of Maryland, College Park, MD 20742 USA}

\author{Or~Katz\orcidlink{0000-0001-7634-1993}}
\affiliation{Duke Quantum Center, Department of Physics and Electrical and Computer Engineering, Duke University, Durham, NC 27701 USA}
\affiliation{School of Applied and Engineering Physics, Cornell University, Ithaca, NY 14853.}

\author{Christopher~Monroe}
\affiliation{Joint Center for Quantum Information and Computer Science, NIST/University of Maryland, College Park, MD 20742 USA}
\affiliation{Joint Quantum Institute, NIST/University of Maryland, College Park, MD 20742 USA}
\affiliation{Duke Quantum Center, Department of Physics and Electrical and Computer Engineering, Duke University, Durham, NC 27701 USA}

\author{Zohreh~Davoudi\orcidlink{0000-0002-7288-2810}} 
\affiliation{Joint Center for Quantum Information and Computer Science, NIST/University of Maryland, College Park, MD 20742 USA}
\affiliation{Maryland Center for Fundamental Physics and Department of Physics, University of Maryland, College Park, MD 20742, USA}

\author{Alexey~V.~Gorshkov}
\affiliation{Joint Center for Quantum Information and Computer Science, NIST/University of Maryland, College Park, MD 20742 USA}
\affiliation{Joint Quantum Institute, NIST/University of Maryland, College Park, MD 20742 USA}

\date{June 8, 2025}
\begin{abstract}

Studying high-energy collisions of composite particles, such as hadrons and nuclei, is an outstanding goal for quantum simulators. However, preparation of hadronic wave packets has posed a significant challenge, due to the complexity of hadrons and the precise structure of wave packets. This has limited demonstrations of hadron scattering on quantum simulators to date. Observations of confinement and composite excitations in quantum spin systems have opened up the possibility to explore scattering dynamics in spin models. In this article, we develop two methods to create entangled spin states corresponding to wave packets of composite particles in analog quantum simulators of Ising spin Hamiltonians. One wave-packet preparation method uses the blockade effect enabled by beyond-nearest-neighbor Ising spin interactions. The other method utilizes a quantum-bus-mediated exchange, such as the native spin-phonon coupling in trapped-ion arrays. With a focus on trapped-ion simulators, we numerically benchmark both methods and show that high-fidelity wave packets can be achieved in near-term experiments. We numerically study scattering of wave packets for experimentally realizable parameters in the Ising model and find inelastic-scattering regimes, corresponding to particle production in the scattering event, with prominent and distinct experimental signals. Our proposal, therefore, demonstrates the potential of observing inelastic scattering in near-term quantum simulators. 
 
\end{abstract} 

\maketitle

\begin{figure*}[t]
    \centering
    \includegraphics[width=2\columnwidth]{  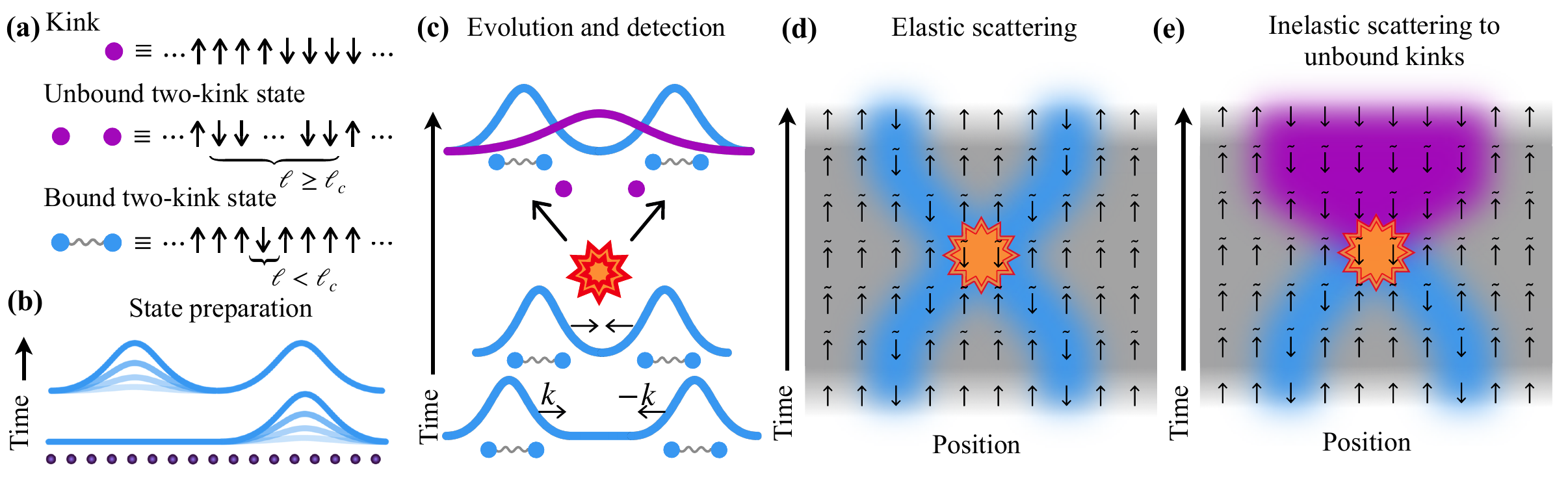}
    
    \caption{\textbf{Scattering of meson wave packets.} 
   (a) Kinks, which are neighboring pairs of anti-aligned spins, can form bound and unbound two-kink states, as determined by the model parameters. We are interested in studying the scattering of bound two-kink states, called mesons, at fixed values of the transverse field $h^z$. 
    Bound two-kink states form stable quasi-particles while unbound two-kink states form unstable quasi-particles that decay into multi-particle states. 
    Two-kink spin configurations at $h^z=0$ can be labeled by the distance $\ell$ between the pair of kinks in the chain. We then define a critical separation $\ell_c$, determined by the model parameters, such that the $h^z = 0$ spin configurations with two kinks separated by $\ell < \ell_c$ ($\ell \geq \ell_c$) are adiabatically connected to bound (unbound) two-kink states at a given $h^z$. Here, an unbound two-kink state is depicted as two isolated purple points and a bound two-kink state (meson) is depicted as blue points joined by a spring.
    (b) An illustration of the creation of localized spin excitations on top of the ground state of the model with $\hat{H}'=0$ [see \cref{eq: Ising Model Hamiltonian}], distributed according to a wave-packet profile. Darker shades of blue correspond to later stages of the evolution. (c) Propagation and collision of meson wave packets resulting in elastic scattering (blue) and inelastic scattering into an unbound two-kink state (purple).  (d) Elastic scattering of two incoming $1$-meson wave packets into two outgoing $1$-meson wave packets. (e) Inelastic scattering between two 1-mesons scattering into two unbound kinks moving away from each other (expanding the purple region). The gradient regions in (d) and (e) from white to gray at early times mark adiabatic ramps from $\hat{H}_0$ to $\hat{H}$. These ramps prepare (dressed) $1$-meson wave packets in the full interacting model from the (bare) wave packets of the free model with only $\hat{H}_0$. The gradient regions back from gray to white at late times mark the reverse ramps that convert dressed wave packets into bare ones. The tilde notation is used to indicate dressed states. 
    }
    \label{fig:MesonIllustrations}
\end{figure*}  

\section{Introduction}
\noindent
A complete understanding of why the strong force in nature confines `color' charges, e.g., quarks and gluons, is still lacking. Decades of 
theoretical and numerical studies have continued to shed light on the mechanism of confinement~\cite{brambilla2014qcd}. Most existing studies examine confinement in static equilibrium settings, with the prominent example being the determination of the static potential between non-dynamical quarks using lattice-gauge-theory methods~\cite{Wilson:1974sk}. These studies establish a linearly rising confining potential as a function of the distance between the charges~\cite{Creutz:1983ev,APE:1987ehd,Bali:1992ru}. They also point to the presence of gluon flux tubes, i.e., strings, between these quarks~\cite{Sommer1987Chromoflux,bali1995observing,haymaker1996distribution,bissey2007gluon}. Nonetheless, confinement and evolution of strings likely also play an important role in non-equilibrium physics of strong interactions. This, however, is much harder to study using current classical numerical methods. To eliminate reliance on predictions based on phenomenological low-dimensional models~\cite{Andersson:1983ia} or on perturbative approaches with limited applicability~\cite{berges2021qcd}, a first-principles approach based on the Standard Model of particle physics is needed. Hamiltonian simulation of real-time dynamics, enabled by quantum simulators, offers a promising route toward this overarching goal~\cite{banuls2020simulating,halimeh2023cold,klco2022standard,bauer2023quantum,Bauer:2023qgm,di2023quantum}. 

Current quantum simulators have not yet reached capabilities needed for studies of the quantum field theories of the Standard Model. It is, nonetheless, important to take full advantage of existing platforms to study non-equilibrium dynamics of models that share salient features of the strong interaction, including confinement, to gain new insights, and to set the stage for more complex simulations in the future. Indeed, analogous confining forces are also present in spin systems, such as the one-dimensional Ising model \cite{mccoy_two-dimensional_1978,DELFINO1996469,kormos_real-time_2017,liu_confined_2019}. Since such spin Hamiltonians have long been engineered and studied in a range of analog quantum simulators~\cite{georgescu2014quantum}, including trapped ions~\cite{RevModPhys.93.025001} and neutral atoms~\cite{henriet2020quantum}, they provide a natural setting to study toy models of confinement, despite the underlying physics behind the confinement being different from that in the Standard Model. In the one-dimensional ferromagnetic Ising spin model, the elementary excitations are kinks, or `domain walls', which are anti-aligned neighboring spins, see \cref{fig:MesonIllustrations}(a). Two kinks can be bound together by a string of anti-aligned spins, to form bound two-kink states, also called mesons. Bound-state spectra~\cite{rutkevich2008energy,kormos_real-time_2017,jurcevic_spectroscopy_2015, kranzl_observation_2022, knaute_relativistic_2022}, string breaking~\cite{pichler_real-time_2016,liu_confined_2019,surace_lattice_2020, tan_domain-wall_2021,verdel_real-time_2020, verdel_dynamical_2023, de2024observation, surace2024stringbreaking,
luo2025quantum}, 
and slow thermalization~\cite{ jurcevic_quasiparticle_2014,robinson2019signatures,lerose2020quasilocalized, lerose_quasilocalized_2019, tan_domain-wall_2021, birnkammer_prethermalization_2022} have been studied, both theoretically and experimentally, in these systems in recent years.

To create non-equilibrium conditions in Ising spin systems exhibiting confinement, we go beyond global quench processes and focus on scattering of individual excitations on top of the interacting vacuum. This is motivated by the fact that much of the strong-interaction dynamics are studied in high-energy particle colliders~\cite{lovato2022long,achenbach2023present}. These experiments often rely on colliding hadrons or nuclei, which are composite (bound) excitations of the elementary quarks and gluons, hence generating a plethora of final-state particles. How non-perturbative confining dynamics lead to these complex inelastic channels through various hadronization and fragmentation mechanisms~\cite{Andersson:1983ia,accardi2009parton,albino2010hadronization} is an intriguing question.
Tensor-network methods have proven a powerful numerical tool to study scattering in one-dimensional models~\cite{van_damme_real-time_2021,milsted_collisions_2022,rigobello_entanglement_2021,belyansky_high-energy_2023,papaefstathiou2024real}, but simulating general scattering problems in quantum field theories has remained out of reach.
Several proposals have emerged recently on how to realize scattering states and processes in simple low-dimensional field theories on digital~\cite{jordan2012quantum,barata2021single,turco2023towards,vary2023simulating,chai2023entanglement,farrell2024quantum,davoudi2024scattering} and analog~\cite{surace_scattering_2021,belyansky_high-energy_2023,su_cold-atom_2024} quantum simulators, but implementations have remained limited~\cite{chai2023entanglement,farrell2024quantum,davoudi2024scattering}. It is, therefore, valuable to study the scattering problem in the simpler spin models, which may provide a more realistic path to large-scale experimental implementations. This still demands preparing composite (bound) states in the form of moving wave packets, which can be made to collide. It will also be important to investigate what types of Ising Hamiltonians and initial states can lead to non-trivial inelastic scattering, going beyond present studies which have found elastic and inelastic scattering in nearest-neighbor models \cite{surace_scattering_2021,karpov_spatiotemporal_2022,milsted_collisions_2022,van_damme_real-time_2021} as well as power-law models, which, nonetheless, 
require non-standard Hamiltonian engineering to generate inelastic processes~\cite{vovrosh_dynamical_2022,wang2024quantum}. Our work sets out to advance the state-of-the-art by addressing these two requirements.

Concretely, we outline an experimental proposal covering the three central stages of scattering: (i) preparation of Gaussian meson wave packets in one-dimensional Ising spin models [see \cref{fig:MesonIllustrations}(b)], (ii) propagation and scattering of the wave packets, and finally (iii) detection of outgoing states [see \cref{fig:MesonIllustrations}(c)].  We propose two techniques to prepare wave packets of bound kinks: one scheme engineers a collective transition to a localized wave packet using beyond nearest-neighbor spin-spin couplings, while the other utilizes a (bosonic) quantum bus to enable excitations needed to prepare the wave packet. An adiabatic ramp is used to evolve the wave packets in the `free' theory to those in the `interacting' theory. Additionally, we study scattering of bound kinks in models with long-range Ising couplings, where kinks are confined, and in models with short-range Ising couplings, which exhibit both bound and unbound kinks. Using numerical simulations of scattering at different energies, we demonstrate elastic scattering in the former, illustrated in \cref{fig:MesonIllustrations}(d), and both elastic and inelastic scattering in the latter for highly energetic wave packets, 
with inelastic scattering 
illustrated in \cref{fig:MesonIllustrations}(e). Importantly, we argue that the inelastic channel can be resolved in near-term analog-simulation experiments. This is established by demonstrating numerical evidence for a non-negligible scattering probability and prominent experimental signature for outgoing free kinks constituting the final state. The remainder of this paper is organized as follows. In \cref{sec:results}, we present our results. In  \cref{sec:discussion}, we present a discussion and outlook for future studies. Finally, in the Appendix, we present some details omitted in the main text.  

\section{Confined and deconfined excitations in the Ising model \label{sec:results}}
\noindent
As a prototypical spin model of confinement and bound excitations~\cite{mccoy_two-dimensional_1978,kormos_real-time_2017,liu_confined_2019}, we consider a one-dimensional quantum spin-$\tfrac{1}{2}$ chain described by the Hamiltonian
\begin{align}
\label{eq: Ising Model Hamiltonian}
    \hat{H} & = \hat{H}_0 + \hat{H}' ,
\end{align}
where
\begin{align}
\label{eq:H0}
    \hat{H}_0 & = - \sum_{i,j>i} J_{ij} \hat{\sigma}^x_i \hat{\sigma}^x_j - h^x \sum_i  \hat{\sigma}^x_i, \\
\label{eq:Hprime}
    \hat{H}' & = - h^z \sum_i  \hat{\sigma}^z_i.
\end{align}
Here, $\hat{\sigma}^x_i$ and $\hat{\sigma}^z_i$ are Pauli matrices acting on the spin at site $i$, with ferromagnetic Ising coupling matrix $J_{ij}$ between spins at site $i$ and site $j$. Additionally, $h^z$ and $h^x$ are the global transverse and longitudinal fields, respectively. We choose the Ising interaction to be along $x$ to match the convention used in trapped-ion-based spin models \cite{RevModPhys.93.025001} and set $\hbar = 1$. 

We consider two types of coupling profiles that can be realized in various quantum simulators; a power-law model, $J_{ij} =  J_0/r_{ij}^\alpha$ with tunable exponent $\alpha > 1$, and an exponentially decaying model, $J_{ij} =  J_0 e^{-\beta(r_{ij}-1)}$ with tunable parameter $\beta> 0$. In both cases, $ J_0 > 0$ is a constant and $r_{ij} = |i-j|$. The limit $\beta\gg1$ or $\alpha \gg 1$ recovers the nearest-neighbor model with uniform coupling $J_0$. Trapped-ion systems can realize either of these models with tunable $0\lesssim\alpha\lesssim3$~\cite{PhysRevLett.92.207901,RevModPhys.93.025001} or $0 < \beta$~\cite{nevado2016hidden,Feng2023,schuckert2023observation,katz2024observing,kim2009entanglement,lee2016floquet}. The power-law model with $\alpha = 6$ (van der Waals interactions) can be realized in neutral atoms encoding the two-dimensional Hilbert space of a spin in a ground state and a Rydberg state~\cite{Schau2012,bluvstein2020controlling,bernien_probing_2017,browaeys2020many}. The power-law model with $\alpha = 3$ (dipolar interactions) can be realized in polar molecules~\cite{Yan2013}, magnetic atoms~\cite{Lahaye2009,chomaz2022dipolar}, and neutral atoms interacting via Rydberg-Rydberg interactions~\cite{Gallagher2008,Gnter2013,ravets2014coherent,browaeys2016experimental}. In systems of electric dipoles such as Rydberg atoms and polar molecules, dipolar interactions of the Ising form can be generated either by applying an electric field to partially polarize the states or by dressing states of opposite parity with a microwave field \cite{gorshkov11a}.

\subsection{Low-energy spectrum and scattering states}
To understand the low-energy excitations of the model in \cref{eq: Ising Model Hamiltonian}, it is useful to first study the limit $\hat{H}'=0$, i.e., vanishing transverse field $h^z$. We refer to this limit as the free theory.
In this limit, all eigenstates are $\hat{\sigma}^x$ eigenstates. The ground state is fully polarized in the 
$x$ direction, satisfying $\langle\hat{\sigma}^x_i\rangle = 1$ for all $i$ when $h^x >0$, and is doubly degenerate, satisfying  $\langle\hat{\sigma}^x_i\hat{\sigma}^x_j\rangle = 1$ for all $i$ and $j$ when $h^x=0$. 
Furthermore, the low-energy excitations of the ground state can be thought as being composed of `kinks' and `anti-kinks', 
\begin{align} 
    \label{eq: kink}
    &\ket{\text{kink}, i + \tfrac{1}{2}}  =  
    \left(\prod_{j\leq i} \hat{\sigma}^{-}_j \right) \ket{ \cdots \uparrow \uparrow \uparrow \ldots }_x, \\ 
    \label{eq: anti-kink}
    &\ket{\text{anti-kink}, i - \tfrac{1}{2}} =  \left(\prod_{j \geq i} \hat{\sigma}^{-}_j \right) \ket{ \cdots \uparrow \uparrow \uparrow \ldots }_x, 
\end{align}
where a kink or anti-kink sits between two anti-aligned spins,
and two-kink states where the kinks are separated by distance $\ell$, 
\begin{equation}
\label{eq:lmeson}
\ket{\ell, i} = \left(\prod_{j=i}^{i+\ell-1} \hat{\sigma}^{-}_j \right)\ket{ \cdots \uparrow \uparrow \uparrow \ldots }_x.
\end{equation}
Here, $\hat{\sigma}^-_j=(\ket{\downarrow}_x \bra{\uparrow}_x)_j$ is the spin lowering operator at site $j$. Eigenspaces of $\hat H_0$ can be broken down into subspaces labeled by the number of kinks $K$ and of spin flips $Q$. States $\ket{\ell, i}$ of two kinks separated by distance $\ell$ correspond to $K = 2$ and $Q = \ell$.
For simplicity, we do not distinguish kink and anti-kink states, i.e., \cref{eq: kink} and \cref{eq: anti-kink}, and refer to all such states with a single spin domain wall as kinks throughout. For an illustration, see \cref{fig:MesonIllustrations}(a).

\begin{figure*}
    \centering
    \includegraphics[scale=0.38]{  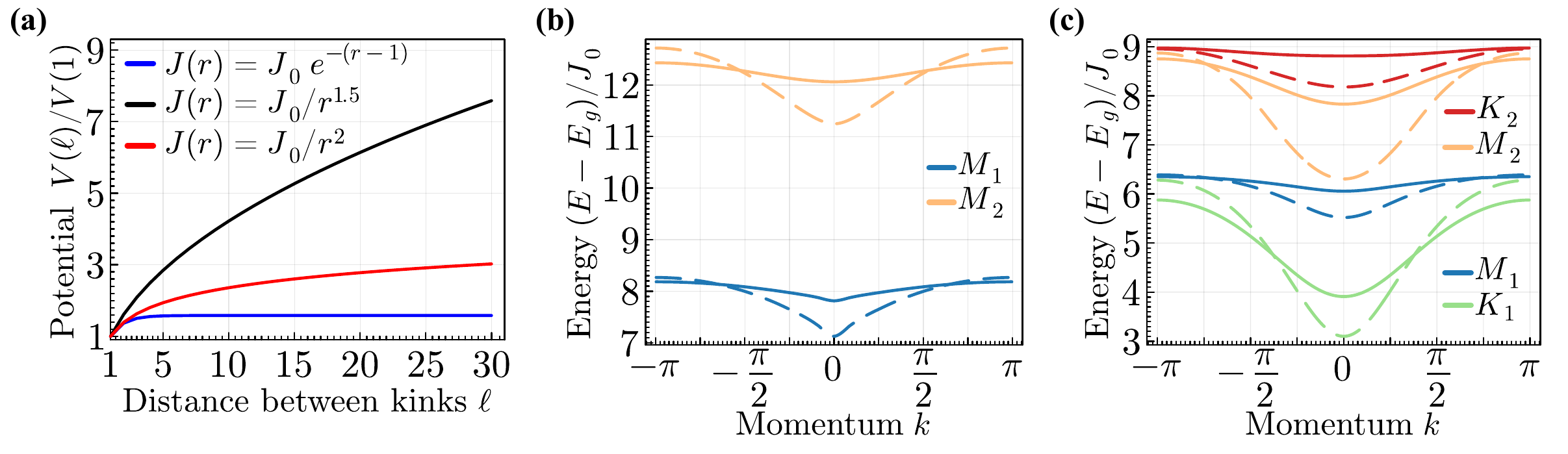}
    \caption{\textbf{Confining potentials and low-energy-excitation bands.} (a) Attractive potential $V(\ell)$ from \cref{eq:V-of-r} between pairs of kinks separated by distance $\ell$ and normalized with respect to $V(1)$ for power-law (red and black curves) and exponentially decaying (blue curve) spin-spin couplings. 
    (b) 1-meson ($M_1$) and 2-meson ($M_2$) Bloch bands above the ground state energy $E_g$ at $h^z=0.8 J_0$ (solid curves)  and $h^z=1.4 J_0$ (dashed curves) for the power-law model ($\alpha=1.5$) computed using exact diagonalization with periodic boundary conditions where convergence was checked as a function of system size. (c) Kink ($K_1$ and $K_2$), 1-meson ($M_1$), and 2-meson ($M_2$) Bloch bands above the ground state energy $E_g$ at $h^z=0.5 J_0$ (solid curves)  and $h^z=0.83 J_0$ (dashed curves) for the exponentially decaying coupling ($\beta=1$) calculated using uniform matrix product states \cite{vanderstraeten_tangent-space_2019,van_damme_real-time_2021,belyansky_high-energy_2023}. This method allows for computing the energy bands of the local (meson) excitations and for the kink excitations. In both (b) and (c), as the transverse field increases, band dispersion increases, and kinks and mesons obtain kinetic energy.}
    \label{fig:Potential-Bands}
\end{figure*}

The energy of an arbitrary $\hat{\sigma}^x$-basis state relative to the fully polarized ground state can be written as a sum of two contributions: the energy required for each flipped spin, which for an infinite chain is given by $m_0 = 2 h^x + 4 \sum_{r=1}^{\infty} J(r)$, and an attractive pairwise potential $-4J(r)$ between each pair of flipped spins separated by a  distance $r$.
Here, we have defined $J(r) \coloneqq J_{ij}$ for $|r_i-r_j| \coloneqq r$. 
The form of the spin-spin interactions, therefore, determines the spectrum of kinks and two-kink states that can exist in the system for a given $h^x$. 
Kinks experience an attractive kink-kink interaction potential $V(\ell)$ shown in \cref{fig:Potential-Bands}(a) and given by
\begin{align}
\label{eq:V-of-r}
    V(\ell) & = \ell m_0 - 4 \sum_{i = 1}^{\ell-1} \sum_{r=1}^{\ell - i}  J(r). 
\end{align}
For systems with $h^x>0$, or for systems with $h^x=0$ and an interaction coupling that decays sufficiently slowly, i.e., power-law decay with $1 < \alpha \leq 2$,
the potential energy $V(\ell)$ of a pair of kinks increases without bound as the distance $\ell$ between the kinks is increased~\cite{liu_confined_2019,lerose_quasilocalized_2019}. For an example, see the black and red curves in \cref{fig:Potential-Bands}(a). This confines all kinks into two-kink bound states labeled by the distance $\ell$ and hence called $\ell$-mesons, shown in \cref{fig:MesonIllustrations}(a) for $\ell = 1$ and described by \cref{eq:lmeson}.
When $h^x = 0$ and the couplings decay more quickly, i.e., exponential decay with $\beta > 0$ or power-law decay with $\alpha > 2$, the kinks experience an interaction potential that saturates at long distances, see blue curve in \cref{fig:Potential-Bands}(a). In this case, as $\ell$-meson energies converge to the finite value $V(\infty)$, 
an infinitesimal transverse field will unbind pairs of kinks separated by an infinite distance.
In the presence of a non-vanishing transverse field, we can define a critical separation $\ell_c$ between two kinks such that $h^z$ endows kinks separated by $\ell \geq \ell_c$ with enough kinetic energy to unbind them~\cite{surace2024stringbreaking}. Thus, the low-energy spectrum consists of a finite number of two-kink bound-states as well as single-kink states below the continuum of unbound kink states~\cite{collura2022discrete}.
Under these circumstances, for a small transverse field, the form given in \cref{eq:lmeson} describes bound two-kink states when $\ell<\ell_c$ and unbound two-kink states when $\ell \geq \ell_c$. Although pairs of kinks are no longer confined for all $\ell$, therefore breaking the analogy with confined quarks, we continue to refer to these two-kink bound states as $\ell$-mesons. Note that, in the nearest-neighbor model, confinement can only occur if $h^x \neq 0$~\cite{mccoy_two-dimensional_1978,DELFINO1996469}.

Let us now consider the model when $\hat{H}' \neq 0$, which we refer to as the interacting theory. Here, we distinguish `bare' kinks and `bare' meson states, i.e., those that are eigenstates of $\hat{H}' = 0$, from `dressed' kinks and `dressed' mesons, i.e., those that are adiabatically connected to their bare analogs, which are dressed by the transverse field $h^z$. 
The low-energy spectrum consists of translationally symmetric meson and kink states with conserved momentum and with energy eigenvalues specified by Bloch bands (i.e., dispersion relation) $E_{n}(k)$. As $h^z$ is adiabatically increased, the kink and meson bands evolve from initially flat bands to bands with dispersion, giving the excitations kinetic energy and a momentum-dependent velocity, see \cref{fig:Potential-Bands}(b) and \ref{fig:Potential-Bands}(c) for the numerically-evaluated lowest-lying Bloch bands of a power-law and an exponentially decaying model, respectively. 

These bands evolve adiabatically unless they intersect higher or lower energy bands. In the models considered here, these intersections occur in the following order: first, the meson bands above the lowest meson band may intersect with the multi-particle continuum, starting with the highest-energy meson bands. 
When an $\ell$-meson band intersects the multi-particle continuum, it indicates that those $\ell$-mesons are no longer stable bound states but unstable states that can decay into multi-particle states. When this occurs, it sets a new critical threshold $\ell_c$ below which the $\ell$-mesons are stable bound states. Finally, the gap between the lowest band and the ground state closes, leading to a phase transition from the ferromagnetic to the paramagnetic phase of the Ising model. 
In this work, we will always remain in the ferromagnetic phase, where we find there is always at least one bound state (meson) that sits below the multi-particle continuum and above the ground state, see \cref{fig:Potential-Bands}(b) and \ref{fig:Potential-Bands}(c). Depending on the values of the parameters, there are possibly several two-kink bound states (mesons)~\cite{liu_confined_2019,collura2022discrete}.  

In this paper, we consider the scattering of two incoming $1$-mesons, the lowest energy two-kink bound states. Open outgoing scattering channels must conserve the energy and momentum of the incoming states. Elastic scattering is always allowed and is composed of the two 1-mesons being transmitted or reflected. Inelastic scattering into new mesons or kinks is kinematically allowed if the bands of the outgoing states, $m$, conserve energy, $E_1(k_1) + E_1(k_2) = 
\sum_{m} E_m(k_m) $, and momentum, $k_1 + k_2 = \sum_m k_m \mod 2\pi$. The prominence of various kinematically allowed outgoing channels is encapsulated by the scattering amplitudes. In this work, the scattering amplitudes are obtained by performing real-time wave-packet collisions and projecting the final states into different meson and kink sectors. 

Future experimental demonstrations of scattering will be implemented on finite chains, which introduces finite-size effects. On a finite chain, kinks and mesons feel the effects of a boundary induced by the absence of the infinite number of spins to the left and right of the finite chain---which becomes increasingly severe as the range of interactions increases~\cite{lerose_quasilocalized_2019}. To mitigate finite-size effects, an effective site-dependent longitudinal field can be introduced, which we call the `pseudo-infinite potential'. This field mimics a potential imposed by an infinite number of fictitious spins frozen in the $\langle \hat{\sigma}_x \rangle = 1$ direction to the left and right of the $N$ dynamical spins (a similar idea has been introduced to study real-time string-breaking dynamics~\cite{surace2024stringbreaking, de2024observation}). While these frozen spins do not fully describe the physics of an infinite chain with all-dynamical spins, this strategy 
alleviates the boundary effects, increasingly so as $h^z$ is decreased, and allows introducing approximate momentum bands of scattering states. Additional details are presented in the \cref{app: pseudo-inf pot}. All numerical scattering simulations presented in this work are performed with this pseudo-infinite potential when applicable.

\section{Experimental scattering proposal
\label{sec:proposal}} 
In order to study the scattering dynamics of bound states in this model, one needs to prepare, scatter, and detect propagating Gaussian meson wave packets. We design our protocol around scattering the lowest-energy bound states, i.e., 1-mesons. We devise strategies to (i-a) prepare two 1-meson Gaussian wave packets in the $\hat{H}_0$ model with opposite momentum on the left and right sides of the chain, (i-b) adiabatically change $\hat{H}_0$ to $\hat{H}$ by ramping the transverse field to prepare wave packets in the full interacting model, (ii) evolve the propagating wave packets under $e^{-i\hat{H}t}$ until the collision is concluded and outgoing particles are sufficiently distant, (iii-a) ramp down the transverse field to return to $\hat{H}_0$, (iii-b) measure and analyze the final states. 

In the $\hat{H}_0$ model, a $1$-meson wave packet is simply a superposition of all single-spin-flip states with a normalized Gaussian amplitude, $\psi^g_i(x_0,k_0)$, centered at $x_0$ with momentum $k_0$, 

\begin{align} 
    \ket{\psi_g(x_0, k_0)} & =  \hat{\Psi}^\dagger_g(x_0,k_0)\ket{ \uparrow}^{\otimes N}_x \\ \label{eq: Undressed gaussian wave packet}
    & = \sum_{i=1}^N \psi^g_i(x_0, k_0 ) \ket{1,i},
\end{align}
with
\begin{align}
    \psi^g_i(x_0, k_0 ) & = \frac{1}{\mathcal{N}} e^{-(x_i-x_0)^2/(2 \Delta_x^2) + i k_0 x_i }.
\end{align}
Here, $ \hat{\Psi}^{\dagger}_g(x_0,k_0) = \sum_{i=1}^N \psi_i^g(x_0, k_0) \hat{\sigma}_i^-$ acts on the all-spin-up state to create a $1$-meson wave packet centered at $x_0$ and $k_0$, $\Delta_x$ is the width of the (position-space) wave packet, and $\mathcal{N}$ is normalization factor such that the wave-packet state is normalized to unity.

In the subsequent sections, we describe two protocols to prepare the state in \cref{eq: Undressed gaussian wave packet} in the $\hat{H}_0$ model. An added benefit of preparing wave packets in the $\hat{H}_0$ model is that the $1$-meson bands have no dispersion, allowing one to prepare wave packets with non-vanishing momenta but with no velocity. This implies that the wave packets do not move during preparation, hence simplifying the protocol. Then, to prepare the wave packet in the full interacting model ($\hat{H}' \neq 0$), the transverse field is increased using an adiabatic ramp $h^z(t)=h^z t/t_r$ for $0\le t\le t_r$, implemented by the evolution operator $\hat{U}_{r}(t_r)$, such that the $1$-meson states are dressed by the transverse field
\footnote{For a slow yet finite adiabatic ramp, the time-evolved state retains a large overlap with the instantaneous ground state and generates only a tiny density of local kink pairs leading to small oscillations in meson number. Our dynamical protocol is not expected to be associated with the visible oscillations that appear for quench protocols in similar confining models \cite{kormos_real-time_2017, tan_domain-wall_2021}.}, 
\begin{align}
\ket*{\tilde{\psi}_g(x_0, k_0)} = \hat{U}_{r}(t_r) \ket{\psi_g(x_0, k_0)}.
\end{align}
Once the wave packets are prepared in the dressed basis, the system can be evolved under the full Hamiltonian $\hat H$ such that the wave packets propagate, collide with each other, and scatter. In order to distinguish the outgoing scattering channels, the transverse field is adiabatically removed so that all final states are once again eigenstates of $\hat{H}_0$. 

The full evolution on the dressed initial states $\ket*{\tilde{\psi}_{\rm init}}$, i.e.,  
\begin{align}
\ket*{\tilde{\psi}_{\rm init}} = \hat{U}_r(t_r) {\hat{\Psi}^{L}_g}^\dagger(x_L, k_L) {\hat{\Psi}^{R}_g}^\dagger(x_R, k_R) \ket{\uparrow}^{\otimes N}_x,
\end{align}
is given by
\begin{align}
\ket{\psi_{\rm final}} = \hat{U}_{r}(t_r)^\dagger e^{-i \hat{H}t} \ket*{\tilde{\psi}_{\rm init}},
\end{align} 
where ${\hat{\Psi}^{L}_g}^\dagger(x_L, k_L)$ only acts on sites $1$ to $N_L  = \lfloor N/2 \rfloor$ and ${\hat{\Psi}^{R}_g}^\dagger(x_R, k_R)$ only acts on sites $N_L + 1$ to $N$. 
Measurements at the end of this protocol provide information about the different final scattering states, which are labeled by the number of kinks $K$ and spin flips $Q$. Gathering statistics of different outgoing scattering channels yields a determination of the scattering $S$-matrix, conveying the probability of a particular channel.

Both of our proposed state-preparation schemes engineer a transition from an easy-to-prepare initial state to the Gaussian wave packet described in \cref{eq: Undressed gaussian wave packet}. Our protocols go beyond the preparation of entangled states invariant under arbitrary permutations of spins, like the W state~\cite{bernien_probing_2017, haeffner_scalable_2005, lin_preparation_2016, zheng_fast_2005, cole_dissipative_2021} or GHZ state~\cite{omran_generation_2019, katz_demonstration_2023, pogorelov_compact_2021, lu2019global}, to many-body entangled states with a Gaussian profile which are required for creating wave packets for scattering.

One of our proposed schemes takes advantage of engineered blockade interactions to generate the wave packet out of an all-spin-up initial state.
The other proposed scheme benefits from quantum-bus-mediated interactions to create wave packets from an all-spin-up state and one excitation in the bus register. We describe these schemes in more detail in the following sections.

\subsection{Blockade wave-packet preparation.} \label{sec: blockade state prep}
Our first state preparation protocol utilizes the Ising spin-spin coupling to excite a single spin flip localized according to a Gaussian distribution. This is done by simultaneously addressing each spin using a specific drive that will be introduced below. Without any spin-spin coupling, each spin would be prepared independently from its neighbors. However, the presence of spin-spin interactions increases the energy of nearby excitations such that only single-spin-flip states are on resonance. Due to its similarity to the blockade in Rydberg arrays~\cite{jaksch_fast_2000, wilk_entanglement_2010, isenhower_demonstration_2010}, we refer to this protocol as blockade wave-packet preparation. 

The protocol can be implemented as follows:
A Gaussian wave packet described in \cref{eq: Undressed gaussian wave packet} is prepared by driving each spin with a site-dependent transverse field $h^z_i(t) = h^z_i \cos( \omega_i t + \phi_i)$, i.e., corresponding to the system Hamiltonian $\hat{H}_0-\sum_i h^z_i(t) \hat{\sigma}_i^z$.  The driving frequency should be set to $\omega_i = E_i - E_0$ where $E_i$ is the energy of the eigenstate $\ket{1, i}$ of the $\hat{H}_0$ Hamiltonian and $E_0$ is the energy of the all-spin-up state. Since the pseudo-infinite potential, described in \cref{app: pseudo-inf pot}, reintroduces the translational invariance of the $1$-mesons on a finite chain, the driving frequency $\omega_i$ is the same for all spins. The transition from the all-spin-up initial state to the wave packet in the $\hat H_0$ model is best illustrated after a transformation into a frame that rotates with $\hat{H}_0$, resulting in the Hamiltonian
\begin{align}
        \label{eq:H blockade rotating frame}
        \hat H_{\rm R}  = & - \frac{1}{2}   \sum_i h^z_i \left( e^{-i \phi_i} \ket{1,i} \bra{\uparrow}^{\otimes N}_x +\text{h.c.}\right) \nonumber \\
        & -  \frac{1}{2} \sum_{i,j} h^z_j \left(  e^{i (\delta_{ij} t - \phi_j)} \hat{\sigma}_j^{-}\ket{1,i}\bra{1,i} + \text{h.c.}  \right) \nonumber \\
        & + \cdots .
\end{align}
Here, $\delta_{ij} = 4 J_{ij}$ is the interaction energy between spins at sites $i$ and $j$. To prepare a Gaussian wave packet, the transverse-field amplitude should be proportional to the desired Gaussian wave-packet amplitude with no momentum, i.e., $h^z_i = \Gamma  J_0 \psi^g_i(x_0, 0)$ with a tunable parameter $\Gamma$, while the site-dependent phase shift should be set to $\phi_i = - k_0 x_i$. Given this choice, the first term describes a resonant transition to the Gaussian wave packet from the all-spin-up state, while the second term describes the closest off-resonant processes to two-spin-flip from single-spin-flip states. These off-resonant transitions are detuned by $\delta_{ij}$. The ellipses denote all other off-resonant contributions.  If driven slowly enough, as explained below, the state prepared at $ J_0 T = \pi/\Gamma$ is the desired wave packet.

\begin{figure*} 
    \centering
    \includegraphics[scale=0.375]{  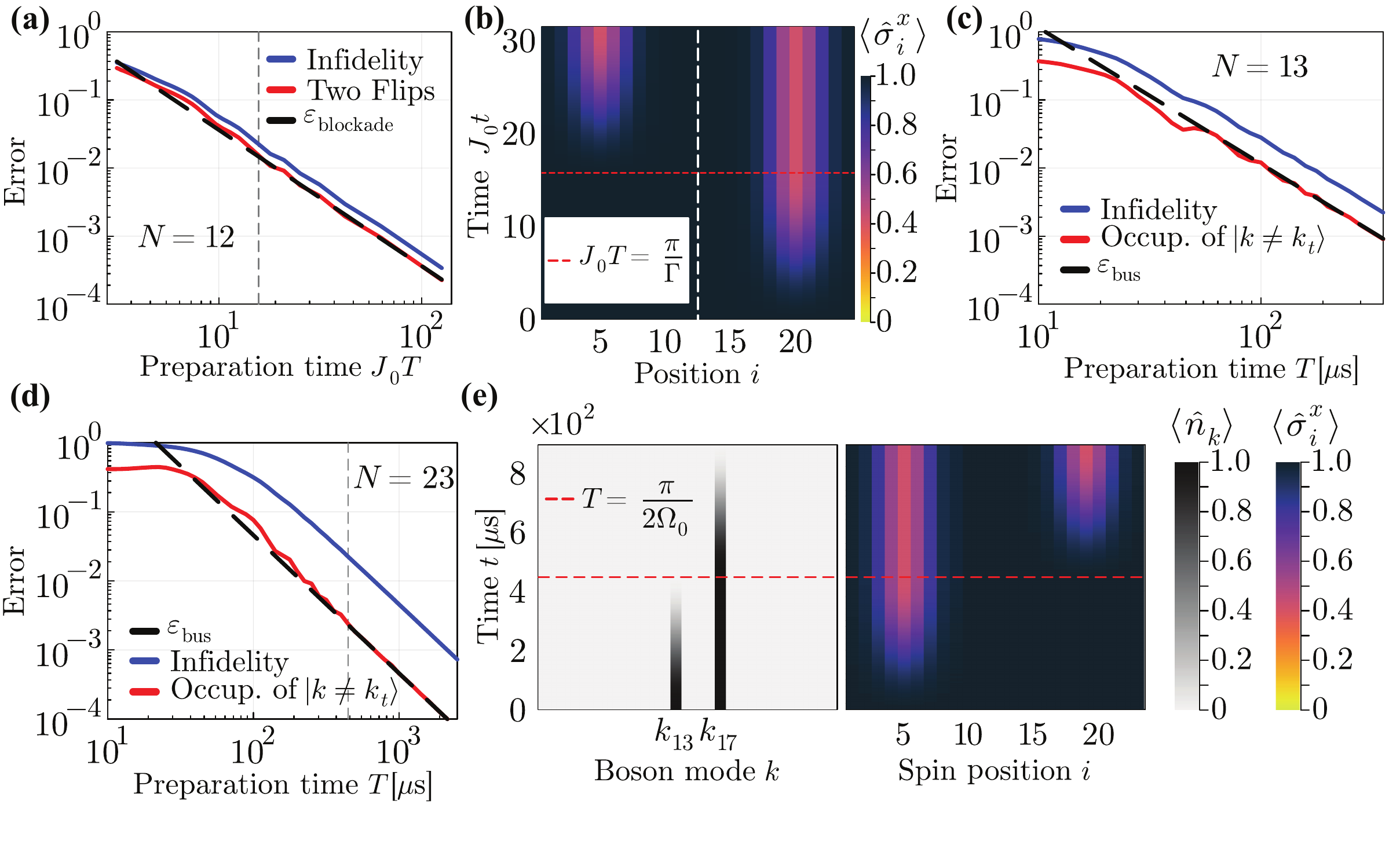}
    \caption{\textbf{Preparation of $1$-meson wave packets.} 
    (a) Blockade state-preparation infidelity and leakage to the two-spin-flip subspace for preparation of a single wave packet centered at $x_0 = 5$ with $k_0 = \pi/2$ on a chain with $N=12$ spins with $\Delta_x = \sqrt{24/(2\pi)}$ and power-law coupling with $\alpha=1.5$. The expected error $\varepsilon_{\rm blockade}$ due to leakage into the two-spin-flip subspace is shown as the black dashed curve. The dotted gray vertical line denotes the preparation time for the simulation in part (b). (b) Preparation of two wave packets, one on the right of the chain (prepared first) and one on the left (prepared second), with a combined final fidelity of $95\%$ for power-law coupling with $\alpha=1.5$ and total preparation time $J_0 T = 31.5$. The white vertical dashed line separates the left and right sides of the chain. The coupling is turned off on one side of the chain when a wave packet is being created on the other side. 
    (c-e) Quantum-bus-mediated state preparation realized using the spin-phonon coupling present in trapped-ion platforms with a given experimental input for $\omega_k$ and $b_{ik}$, as outlined in \cref{app: state preparation}, and $\Omega_0=\pi/(2T)$ MHz.
    Preparation infidelity and occupation of the non-target phonon modes as a function of preparation time $T$ reported in microseconds for the preparation of a single wave packet at $x_0 = 5$ with $k_0 = \pi/2$ in a (c) $N = 13$ ion chain and (d) $N =23$ ion chain. 
    In (c) and (d), the expected population of the non-target phonon modes $\varepsilon_{\rm bus}$ is shown as the black dashed curve. The dotted gray vertical line in (d) denotes the preparation time for the simulation in part (e). 
    (e) Preparation of two wave packets with $98\%$ combined fidelity using the quantum-bus-mediated scheme for $N = 23$ and total preparation time $T = 893 \mu \rm{s}$. 
    Starting with one quantum initialized in two different phonon modes (left heatmap) and spins in the all-spin-up state (right heatmap), the left wave packet is prepared by coupling spins on the left of the chain to the $13^{\rm th}$ phonon mode; the same is done for the right wave packet but with the $17^{\rm th}$ phonon mode. 
    The horizontal red dashed line in (b) and (e) denotes the end of the first wave-packet preparation. }
    \label{fig:state-prep}
\end{figure*}

The leading-order error occurs when the effect of the second term in \cref{eq:H blockade rotating frame} is significant. Undesired transitions from the Gaussian wave packet to states with two spin flips occur when the transition matrix element is large compared to the interaction energy. This transition error has the approximate form $\varepsilon_{\rm blockade} \coloneqq \sum_{i,j>i} \epsilon_{ij}$ where $\epsilon_{ij} =T^{-2} \abs{ \pi \psi_i^g(x_0,k_0) \psi_j^g(x_0,k_0) / \delta_{ij}}^2$, see \cref{app: state preparation}. For far-away interacting spins, this error is small because the wave packet does not have simultaneous support on sites $i$ and $j$, while for neighboring spins, this error reduces to the condition $\Gamma \ll 4$ which limits the magnitude of the driving field $h^z$ to small amplitudes during the blockade wave-packet preparation.
The performance of this scheme for a single Gaussian wave packet over $N=12$ sites with $\Delta_x = \sqrt{24/(2\pi)}$ and power-law coupling with $\alpha=1.5$ is shown in \cref{fig:state-prep}(a). In order to reach infidelities lower than 0.1 (0.01), preparation times greater than $7 J_0$ ($25 J_0$) are required. Additional details can be found in \cref{app: state preparation}. The Appendix also includes a discussion on generating larger, more energetic $\ell$-meson wave-packets.

\subsection{Quantum-bus-mediated wave-packet preparation.
\label{sec: bus mediated state prep}}
A second state-preparation scheme can be implemented by utilizing a quantum bus with controlled coupling to the spins within the wave packet. The idea is to transfer an excitation in the quantum-bus register to the spin registers such that the excitation is spatially distributed according to a desired wave-packet profile. The exact amplitude can be controlled by tuning the strength of the coupling to each spin in the wave packet. The quantum bus can be either a bosonic mode or a spin. Examples with bosonic-mode buses include ion spins coupled to collective phonon modes in trapped-ion platforms~\cite{PhysRevLett.92.207901,RevModPhys.93.025001,katz_programmable_2023}, Rydberg tweezer arrays coupled to optical cavities~\cite{PRXQuantum.3.010344}, as well as transmon qubits coupled to microwave resonators in circuit QED~\cite{PhysRevA.82.043811}. An example where the bus is a spin is a Rydberg-atom array where interactions are tuned such that a chosen bus atom interacts with the other atoms, while the other atoms do not interact with each other~\cite{saffman09,young21a}.

In this work, we consider systems where the quantum bus is a bosonic mode. However, for a spin bus, the treatment is identical, with an additional simplification that it is trivial to prepare a single excitation in the bus.
An example of a time-dependent Hamiltonian describing the coupling between a bosonic quantum bus and spins is given by the anti-Jaynes-Cummings Hamiltonian,
\begin{align} 
\label{eq: spin phonon coupling}
    \hat{H}(t) = \sum_{i,k} \left( A_{ik}  e^{i \delta_k t} \hat \sigma^-_i \hat a_k + A^*_{ik}  e^{-i \delta_k t} \hat \sigma^+_i \hat a_k^\dagger \right).
\end{align}
Here, $\hat{\sigma}^+_i = (\ket{\uparrow}_x\bra{\downarrow}_x)_i$ is the spin raising operator at site $i$, $\hat{\sigma}^-_i = (\ket{\downarrow}_x \bra{\uparrow}_x)_i$ is the spin lowering operator at site $i$, $\hat a_k^\dagger $ ($\hat a_k$) is the boson creation (annihilation) operator for mode $k$, $A_{ik}$ are the site- and mode-dependent amplitudes, and $\delta_k = \omega_k - \nu$ are the detunings from the boson mode frequency $\omega_k$ with $\nu$ being the drive frequency. Note that site- and mode-dependent amplitudes can be realized in, e.g., trapped-ion systems by driving the blue-sideband transitions with $A_{ik} = \eta_k b_{ik} \Omega_i/2$ where $\eta_k$ is the Lamb-Dicke parameter of mode $k$, $b_{ik}$ are the site-dependent orthonormal mode-participation matrix elements of the collective phonon modes, and $\Omega_i$ is a site-dependent Rabi frequency~\cite{wineland1998experimental, RevModPhys.93.025001}. Note that we could have chosen the Ising-Hamiltonian parameters such that the ground state in absence of $H'$ is an all-spin-down state. The scattering could then be performed with $1$-mesons defined as $\ket{1,i} \coloneqq \hat{\sigma}^+_i \ket{\downarrow}_x^{\otimes N}$. In this case, the Jaynes-Cummings Hamiltonian, which has terms such as $\hat{\sigma}^+_i \hat{a}_k$, can be used to perform the quantum-bus-mediated preparation.  

The idea of this scheme is to first start from a state initialized with all spins up and no occupied bosonic modes, then create an excitation in the target boson mode and, finally, use the evolution under \cref{eq: spin phonon coupling} to transfer the bosonic excitation to the chain. To prepare an excitation in the target boson mode, first the $j^{\rm th}$ spin is flipped down, then, by evolving under \cref{eq: spin phonon coupling}, this down spin at site $j$ is flipped back up while the excitation from the spin register is transferred to the target boson mode. Once this state is prepared, the bosonic excitation can be transferred back to the spin register but controlled such that the spin excitation is distributed according to a wave-packet profile. We program the desired wave-packet profile into the time evolution of the target boson mode $k_t$ by setting the amplitude to 
\begin{equation}
    A_{ik} = \Omega_{0} B_{ik} \psi^g_i(x_0, k_0) 
\end{equation}
such that $B_{i k_t} = 1$. Here, $\Omega_0$ is a tunable Rabi frequency. In general, $B_{ik}$ contains information about the coupling strength between the $k^{\rm th}$ boson mode and each spin. Now, if the system is driven at the target-mode frequency, i.e., $\nu = \omega_{k_t}$, one arrives at the Hamiltonian
\begin{align}
    \hat{H}(t) = & \, \Omega_0 \sum_{i} \left( \psi^g_i(x_0, k_0) \hat \sigma^-_i \hat a_{k_t} 
    + \rm{h.c.} \right) \nonumber \\
    & + \Omega_0 
     \sum_{i, k\neq k_t} \left( \psi^g_i(x_0, k_0) B_{ik} e^{i \delta_{k} t}\hat \sigma^-_i \hat a_{k} 
    + \rm{h.c.} \right).
\label{eq:H-bus-mediated}
\end{align}
Here, the first term describes the resonant transfer of an excitation from the target boson mode to a spin excitation distributed according to the Gaussian wave packet when acted on an all-spin-up state. The second term describes the off-resonant transfer between the rest of the boson modes and spin excitations. Starting from an initial state in which the target mode $k_t$ is occupied by one quantum and evolving for $T = \pi/(2 \Omega_0)$ results in the annihilation of the occupied target mode and excitation of a single spin flip in the spin register distributed in the shape of the Gaussian wave packet. 
If the spin-boson coupling cannot be individually controlled in a given experimental platform, then this scheme can be utilized to create localized, but not Gaussian, wave packets. In this case, the target modes whose spin-phonon coupling profile is similar to the desired wave packet should be chosen. An example of such localized modes for the site-dependent mode-participation matrix $b_{ik}$ in trapped-ion platforms is shown in \cref{fig: spin-phonon details}.

\begin{figure*}[t]
    \includegraphics[width=2\columnwidth]{  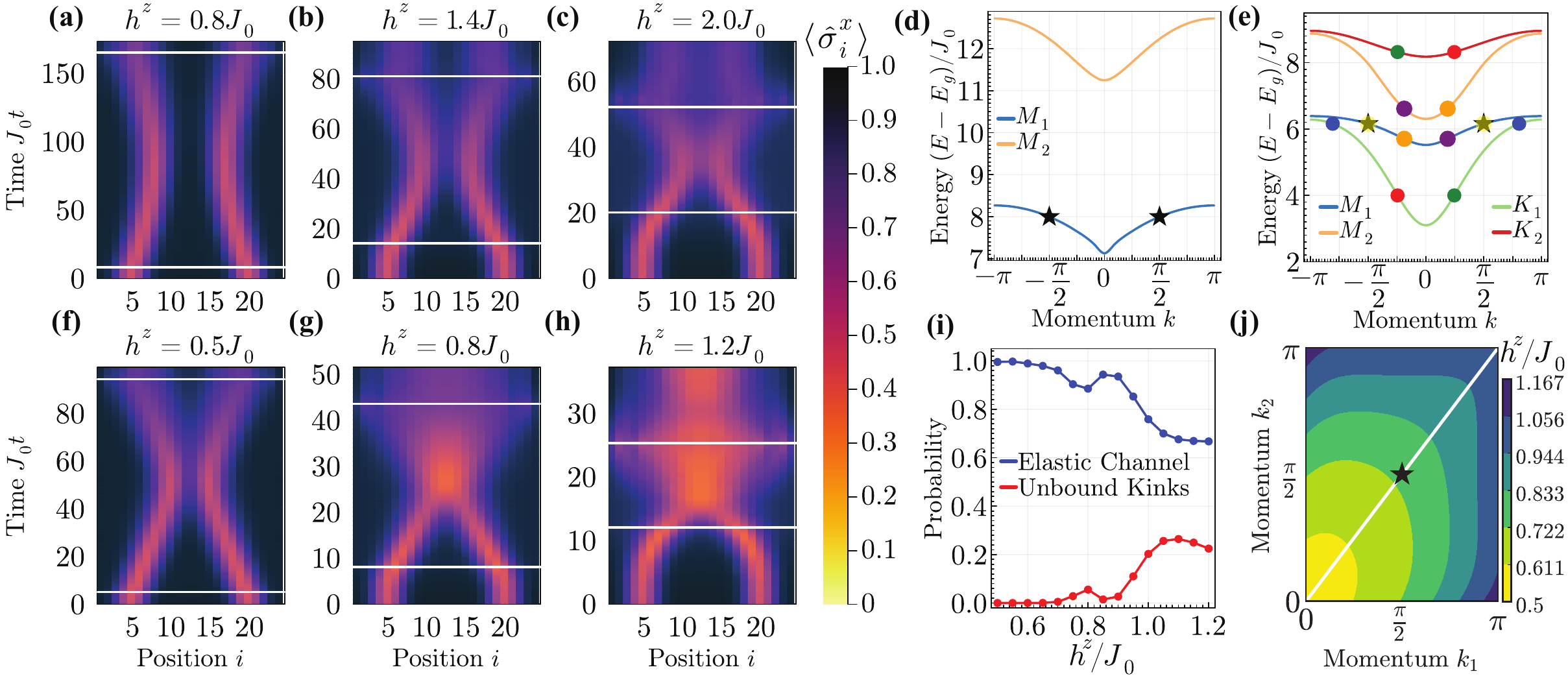}
     \caption{\textbf{Numerical simulation of scattering.} 
        Panels (a-c,f-h) show the expectation value of $\langle \hat{\sigma}^x_i \rangle$ as a function of time for the evolution and scattering of two $1$-mesons for (a-c) the power-law model with $\alpha = 1.5$ and (f-h) the exponentially decaying model with $\beta = 1$. All initial states of the scattering simulations start with a perfect initial two-wave-packet state with $x_0 = 5, k_0 = \pi/2$ for the left wave packet and $x_0 = 20, k_0 = -\pi/2$ for the right wave packet, while the wave-packet width is set to $\Delta_x=\sqrt{N/(2\pi)}$.  The first horizontal white line indicates the end of the linear adiabatic ramp $U_{r}(t_r)$ which turns on $h^z$, while the second horizontal white line indicates the beginning of the linear adiabatic ramp $U_{r}(t_r)^\dagger$ which turns off $h^z$.  (d) In the power-law model for $\alpha=1.5$ and $h^z = 1.4  J_0$, only elastic scattering of 1-mesons with momenta $(\pi/2, -\pi/2)$ is allowed by conservation of energy and momentum, marked by black stars. Here, $M_1$ and $M_2$ denote the two lowest-lying meson bands. (e) By contrast, in the exponentially decaying model for $\beta=1$ and $h^z = 0.83  J_0$, several inelastic outgoing channels are kinematically-allowed: the incoming mesons (marked by black stars) can scatter into mesons of different types or into a pair of unbound kinks.  Here, $M_1$ and $M_2$ denote the two lowest-lying meson bands, while $K_1$ and $K_2$ are the two lowest-lying kink bands. (i) In the exponentially decaying model, the probabilities are plotted as a function of $h^z$ for two mesons scattering into a pair of unbound kinks or elastically into two mesons. Unbound kinks are observed starting at $h^z \approx 0.75  J_0$ and higher. (j) Region of kinematically allowed scattering from mesons into unbound kinks for the exponentially decaying model ($\beta=1$) when the input meson momenta $k_1$ (traveling to the right) and $k_2$ (traveling to the left) are varied. The color corresponds to the value of $h^z/ J_0$ at which the unbound kink channel becomes kinematically allowed for the numerically computed values. For slow moving mesons near $k=0$, the unbound kink channel is kinematically allowed for $h^z \gtrsim 0.5  J_0$. At the momenta $(\pi/2, \pi/2)$, marked by a black star and used in the simulations in panels (a-c, f-i), the unbound kink channel is allowed for $h^z \gtrsim 0.722  J_0$, consistent with the initial rise from zero of the inelastic scattering probability in (d). 
     }
    \label{fig:scattering}
\end{figure*}

Undesired transitions arise when excitations in the spin register transfer to phonon modes other than the target mode. This leakage error occurs when the off-resonant terms in \cref{eq:H-bus-mediated} cannot be neglected,
and has the approximate form $ \varepsilon_{\rm bus} \coloneqq \sum_{k \neq k_t} \epsilon_k $ where $\epsilon_k = T^{-2} \abs{ \frac{\pi }{ 2 \delta_k} \sum_{i=1}^N \abs{ \psi^g_i(x_0, k_0) }^2  B_{ik}  }^2 $, see \cref{app: state preparation}. 
In general, the target mode should be chosen such that it is strongly coupled to the spins involved in the wave packet. For example, in trapped-ion systems, $B_{ik} = \tfrac{ \eta_k b_{ik}}{\eta_{k_t} b_{ik_t}}$, and the target phonon mode is chosen such that $b_{ik_t}$ has maximal support on the wave-packet amplitudes. The performance of this scheme can be benchmarked using platform-specific parameters for trapped ions. Figures~\ref{fig:state-prep}(c) and \ref{fig:state-prep}(d) show the preparation performance of a single wave packet in a chain of $N=13$ and $23$ ions, respectively. Given the experimental parameters chosen, preparing a wave packet with infidelity lower than $0.1$ ($0.01$) requires preparation times of approximately $50 \mu\rm{s}~(180 \mu\rm{s})$ for $N = 13$ and $220 \mu\rm{s}~(700 \mu\rm{s})$ for $N = 23$.  See the \cref{app: state preparation} for additional details on the trapped-ion implementation of the quantum-bus-mediated protocol. 

Two wave packets can be prepared in the chain as follows. For the blockade preparation scheme, controlling the spin-spin coupling $J_{ij}$ to only have support on the left or right side of the chain allows the left and right wave packets to be prepared sequentially without additional sources of error. During the preparation of each wave packet, the pseudo-infinite potential should be chosen such that it corresponds to the smaller half-chain system where $J_{ij}$ is turned on. We note that the spins are subject to reduced noise when spin-spin couplings are turned off. For large chains, wave packets can be prepared sufficiently far away from each other such that they do not interact and can, therefore, be prepared in parallel. Recall that the wave packets do not move during preparation, so they remain separated as they are prepared. 

To prepare two wave packets using the quantum-bus-mediated scheme, one first couples the spins on the left of the chain to one target mode, $k_t^L$, and then couples the spins on the right side of the chain to another target mode, $k_t^R$. The above protocol can be performed sequentially with $A_{ik} \neq 0$ for the sites on which the desired wave packet has support. The two modes should ideally be chosen to maximize the support on the spins to be flipped; however, the same mode can also be used for both wave packets provided it is re-initialized with a single excitation after creating the first wave packet. Parallel preparation using the bus-mediated scheme is also feasible if experimental control allows driving the spins on the left side resonantly with mode $k_t^L$ and the spins on the right side resonantly with mode $k_t^R$. This way, one can effectively independently dial in two independent sets of site-dependent Rabi frequencies $\Omega_{i,k_t^L}$ and $\Omega_{i,k_t^R}$ for the two resonantly driven modes. Parallel preparation could potentially halve the preparation time thus reducing the overall time of the scattering simulation.   

Sequential preparation of two wave packets is studied numerically and the results are shown in \cref{fig:state-prep}(b) for the blockade scheme with $N = 24$ and $J_0 T = 31.5$  and \cref{fig:state-prep}(e) for the bus-mediated scheme with N = 23 with $T = 893 \mu\rm{s}$. 
To compare both schemes in trapped-ion platforms, we note that $J_0$ is of order $\approx 1~\rm{KHz}$ while $\Omega_0$ 
can be 
interpreted as the transition strength when addressing the first sideband and is typically 
of order $\lesssim  \eta_k b_{ik} \times 1~\rm{MHz} \approx 0.01~MHz$. Therefore, in trapped-ion systems $\Omega_0 \gg J_0$ and wave packets are prepared much faster using the quantum-bus-mediated scheme.

\section{Numerical scattering simulations}
Suppose that the wave packets are generated with high fidelities, via one of the methods described earlier in this section. The scattering experiment can then be performed, as outlined previously, to probe non-trivial dynamics of post-collision processes. To develop a theoretical expectation for such dynamics, we perform a numerical simulation by starting with perfect wave packets and studying their time evolution. We demonstrate elastic scattering in the model with power-law coupling, which exhibits only two-kink bound states ($\ell$-mesons)---and no unbound kinks, at low energies (considering the chosen initial state). Additionally, we demonstrate both elastic and inelastic scattering in the model with exponentially decaying coupling, which exhibits both two-kink bound states and unbound kinks in the relevant energy range. Scattering simulations were performed using a Krylov time-evolution method. 

Starting with $1$-meson wave packets as described in \cref{eq: Undressed gaussian wave packet}, we turn on the transverse field using a linear ramp for $t_r = 10 /h^z$ in order to approach the Hamiltonian $\hat{H}$ starting from $\hat{H}_0$. We then evolve in time such that the meson wave packets propagate and scatter. The transverse field is then removed using an inverted linear ramp to return to $\hat{H}_0$. The state is finally projected into different scattering channels as defined by the number of kinks, $K$, and spin flips, $Q$. The elastic channel is the set of bit strings with four kinks and two spin flips, since such a channel is characterized by two 1-mesons, each with two kinks on either side of a single spin flip. Inelastic scattering is detected by support in any channel outside of the elastic channel. Inelastic scattering in the form of unbound kinks is described by the set of bitstrings with two kinks and any number of spin flips greater than two. We note that while we do not measure the momentum of the outgoing particles, if one had access to large chains, the outgoing momentum could be obtained by measuring the velocity of the outgoing states \cite{belyansky_high-energy_2023, surace_scattering_2021}. Varying $h^z$ allows for the exploration of scattering in different energy regimes. 

For scattering in the power-law model with $\alpha=1.5$, as shown in \cref{fig:scattering}(a-c), only elastic scattering is detected in the range of $h^z$ values considered. For the exponentially decaying model with $\beta=1$, we observe both elastic and inelastic scattering in \cref{fig:scattering}(f-h) as $h^z$ increases. The region of flipped magnetization that appears between the outgoing particles in \cref{fig:scattering}(g) and \ref{fig:scattering}(h) is due to an inelastic scattering channel composed of a pair of unbound kinks. When individual shots are measured in the final state, this appears as a domain of flipped spins, whose length grows in time as the kinks fly away from each other. When this flipped domain has grown to a long length, it will be simple to distinguish it from other output states without unbound kinks. For values of $h^z>1$, we observe up to $25\%$ probability for such an unbound-kinks scattering channel, see \cref{fig:scattering}(i). Given this sizable probability and the distinct signature of the associated measurement, this proposal provides an opportunity to detect inelastic scattering in near-term spin quantum simulators.

The existence of the unbound kink channel in the exponentially decaying model is supported by an analysis of the kinematically allowed scattering channels. In \cref{fig:scattering}(e), we show the lowest energy bands of the exponentially decaying model with $\beta=1$ and $h^z = 0.83  J_0$, consisting of two kinks and two mesons. These bands are computed using uniform matrix product states~\cite{vanderstraeten_tangent-space_2019} with the quasi-particle ansatz of Ref.~\cite{PhysRevB.85.100408}. With these numerically determined bands, one can find sets of outgoing particles with total energy and momentum that match those of the incoming particles. For the case of two 1-mesons with momenta $\pm \pi/2$, the resulting sets of outgoing particles are shown in \cref{fig:scattering}(e) as matching pairs of colored circles. Despite the existence of several kinematically allowed inelastic scattering channels, our simulations show that only one such channel has significant probability in the output state for the parameter range we consider. This channel consists of an unbound pair of kinks, each of which belongs to the lowest kink band. For incoming mesons with momenta $\pm \pi/2$, this unbound-kink channel opens up at $h^z \approx 0.72  J_0$, indicated by the black star in \cref{fig:scattering}(j). After the channel becomes open, the unbound-kinks scattering probability shown in \cref{fig:scattering}(i) increases to a peak, then decreases, then increases again. This non-monotonic behavior is dynamically generated in the interacting model and involves $h^z$ values that are inaccessible to a perturbative analysis.

\section{Discussion
\label{sec:discussion}}
\noindent
An exciting promise of quantum simulators is to study real-time dynamics of scattering processes in nuclear and high-energy physics rooted in the fundamental theory of subatomic constituents, i.e., the Standard Model. In order to realize this promise, concrete experimental proposals are needed to prepare complex initial scattering states such as hadronic wave packets. As analog quantum simulations based on the Standard-Model Hamiltonians are not yet feasible, in this work we turned to simpler Ising spin models in one spatial dimension, which exhibit some of the salient features of the theory of the strong force in nature, including confinement of elementary excitations into composite bound states, namely hadrons. We developed detailed experimental proposals to prepare and scatter meson wave packets in such models with both power-law and exponentially decaying spin-spin couplings, and demonstrated well-resolved inelastic scattering to unbound states in the latter case. We further investigated the range of the transverse-field strength for which the inelastic channel becomes sufficiently prominent to be resolved in future experiments. 

Beyond the ability to simulate a one-dimensional Ising Hamiltonian, there are a few additional experimental capabilities that are required to execute the proposal outlined in this work. If an auxiliary bosonic degree of freedom is coupled to the spins, one may implement our quantum-bus-mediated scheme, provided that the spin-boson coupling can be controlled individually at each site.
If only spin-spin interactions can be controlled, one may use our blockade scheme, provided that site-dependent control of a driven transverse field is available. 
Alternatively, in Rydberg arrays, one can use a Rydberg blockade to create bare meson wave packets.
Additionally, experiments need to implement adiabatic ramps of the transverse field to prepare the wave packets in the full interacting model. 
Control over site-dependent longitudinal fields permits the implementation of the pseudo-infinite potential, which reduces the qubit overhead while mitigating the effects of boundaries. Finally, site-resolved spin measurements are required to access final-state probabilities and analyze the scattering channels. As an outlook, our protocols can be further combined with entanglement spectroscopy tools~\cite{kokail2021entanglement,kokail2021quantum,joshi2023exploring,mueller2023quantum}, provided availability of single- and multi-qubit operations, to analyze entanglement structure of the final state and study net entanglement generation in the collision events. Additionally, some of our protocols, such as our wave-packet preparation schemes, may be relevant to meson scattering in other models exhibiting confinement, such as the Ising-Dicke Hamiltonian \cite{rohn2020ising,puel2024confined}.  

Importantly, studying scattering processes using the protocols of this work involves long simulation times, given the requirement of adiabatic evolution of the free isolated wave packets to interacting colliding wave packets, and a reverse adiabatic evolution to recover well-separated free wave packets long after the collision. Depending on the wave-packet preparation procedure and the characteristics of the chosen analog simulator, the entire process may be too long to fit the coherence time of present-day analog simulators. 
Based on the numerical simulations of this work adopted to trapped-ion systems, we find that the blockade state preparation requires $J_0 T = 31.5$ to achieve $95\%$ fidelity for creating two wave packets, the bus-mediated state preparation requires $890 \mu\rm{s}$ to achieve $98\%$ fidelity for creating two wave packets, and the rest of scattering (i.e., the two linear adiabatic ramps of the transverse field and evolution with the full Hamiltonian) takes $J_0T = 70 - 160$ in the power-law model and $J_0 T = 50 - 100$ in the exponentially decaying model, where the ranges correspond to a range of $h^z$ values studied in this work.

For state-of-the-art trapped-ion simulators, currently accessible simulation times are $J_0 t = 10 - 20$ \cite{Feng2023,joshi2022observing}. Given these simulation times, additional fine tunings and improvements to the current protocols may, therefore, 
be needed to enable the first quantum simulation of hadron scattering in an analog quantum simulator. For example, the adiabatic ramp can be optimized, complex laser-pulse-shaping methods can be employed to improve the bus-mediated state preparation, and parallel implementation of the two wave packets in longer chains can be considered to halve the time duration of the state-preparation step. 
Last but not least, numerical simulations incorporating inexact wave packets, realistic hardware-noise models, and specific experimental details may be required to identify robust experimental signatures of the final-state scattering channels.
These studies are left to future work. Our proposal, nonetheless, brings us closer to realizing hadronic scattering in present-day and future analog quantum simulators.
\vspace{0.2 cm}

\noindent
\textbf{Acknowledgments.} We thank Jeffery Yu for a discussion about preparing wave packets using dispersive quantum-non-demolition coupling.
This material is based upon work supported by the U.S.\ Department of Energy (DOE), Office of Science, National Quantum Information Science Research Centers, Quantum Systems Accelerator. Additional support is acknowledged from the following agencies.
 E.R.B.~acknowledges support from the DOE, Office of Science, Office of Advanced Scientific Computing Research (ASCR), Computational Science Graduate Fellowship (award no.~DE-SC0023112).
E.R.B., B.W., A.S., R.B., and A.V.G.~were supported in part by the NSF STAQ program, NSF QLCI (award no.~OMA-2120757), AFOSR MURI, AFOSR, DOE ASCR Quantum Testbed Pathfinder program (awards no.~DE-SC0019040 and DE-SC0024220), DOE ASCR Accelerated Research in Quantum Computing program (awards No.~DE-SC0020312 and No.~DE-SC0025341), and DARPA SAVaNT ADVENT.  
F.M.S.~acknowledges support provided by the DOE, Office of Science, Office of ASCR (award no. DE-SC0020290), by Amazon Web Services, AWS Quantum Program, and by the DOE QuantISED program through the theory consortium ``Intersections of QIS and Theoretical Particle Physics'' at Fermilab.
Z.D.~acknowledges funding by the DOE, Office of Science,
Early Career Award (award no.~DESC0020271), and by the Department of Physics, Maryland Center for Fundamental Physics, and the College of Computer, Mathematical, and Natural Sciences at the University of Maryland, College Park. She is further grateful for the hospitality of Perimeter Institute where part of this work was carried out. Research at Perimeter Institute is supported in part by the Government of Canada through the Department of Innovation, Science, and Economic Development, and by the Province of Ontario through the Ministry of Colleges and Universities. Z.D. was also supported in part by the Simons Foundation through the Simons Foundation Emmy Noether Fellows Program at Perimeter Institute.

\appendix

\section{Pseudo-infinite potential} \label{app: pseudo-inf pot}
\begin{figure}
    \centering
    \includegraphics[scale=0.36]{ 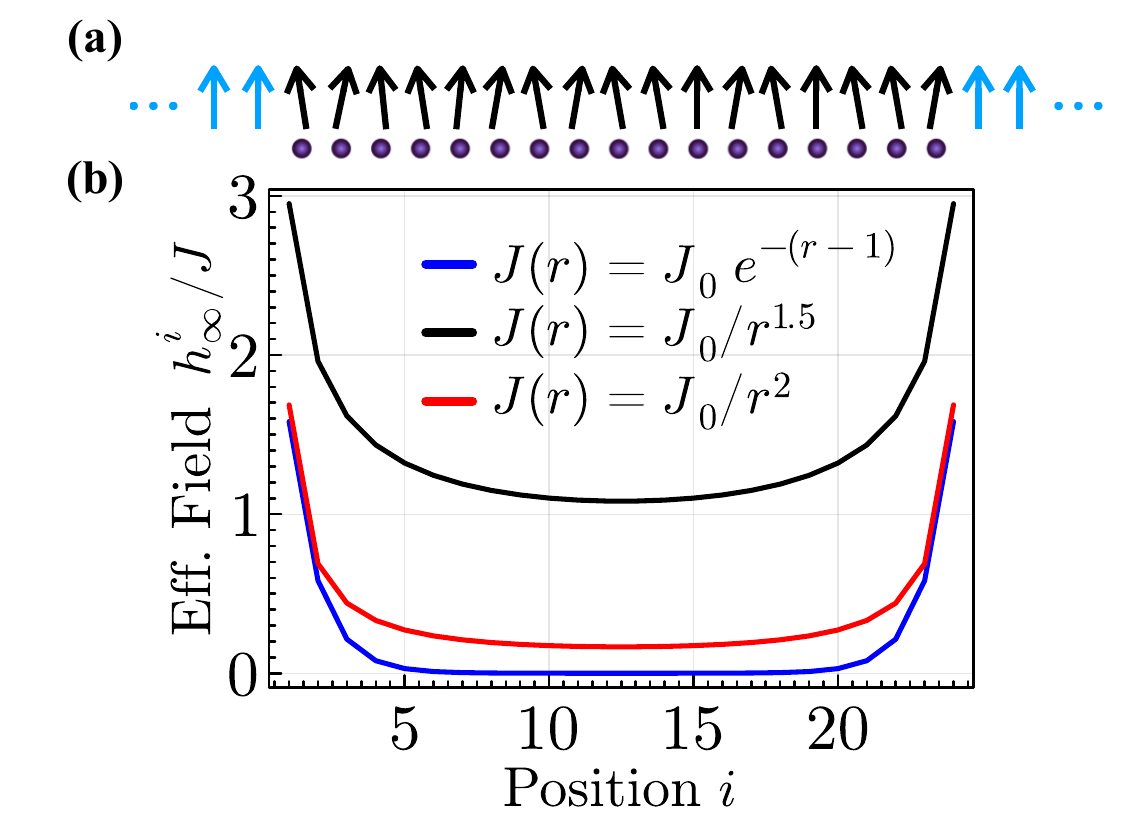}
    \caption{\textbf{Pseudo-infinite potential.} (a) Illustration of fictitious frozen spins (blue) to the left and right of the dynamical
    spins (black) whose effect is taken into account using a pseudo-infinite potential given by \cref{eq: pseudo-inf potential} and shown in (b). The form of the effective potential which mimics these fictitious frozen spins for power-law spin-spin interactions is plotted as red and black curves and for an exponentially decaying coupling as the blue curve. The effective potential is more (less) significant for long-(short-)range spin-spin interactions.}
    \label{fig: pseudo-inf potential}
\end{figure}

In this section, we discuss the explicit form of the pseudo-infinite potential discussed in \cref{sec:results}. 
In order to mitigate boundary effects in a finite chain, we introduce a site-dependent longitudinal field $h^x_i=h^\infty_i$, which we call the pseudo-infinite potential, that mimics a potential imposed by an infinite number of fictitious spins frozen in the $\langle \hat{\sigma}_x \rangle = 1$ direction to the left and right of the $N$ dynamical spins, as shown in \cref{fig: pseudo-inf potential}(a). A similar idea has been introduced to study real-time string-breaking dynamics in Refs.~\cite{surace2024stringbreaking, de2024observation}. 
This pseudo-infinite potential is given by
\begin{equation} 
\label{eq: pseudo-inf potential}
    h_i^\infty = \sum_{n = -\infty}^{0} J_{ni} + \sum_{n = N+1}^\infty J_{in},
\end{equation}
and is plotted in \cref{fig: pseudo-inf potential}(b) for power-law and exponentially decaying spin-spin interactions. The effect of this potential is more significant for long-range interactions and is less important for short-range interactions. In addition to reintroducing (approximate) translational invariance of the meson and kink states, this pseudo-infinite potential also allows meson wave packets to propagate closer to the chain's edge without getting too distorted thus reducing the required number of spins for an experiment. Since the fictitious spins are frozen and not dynamic, the pseudo-infinite potential does not entirely alleviate finite-size effects. To implement this pseudo-infinite potential in an experimental device, one may e.g., find the $\alpha$ or $\beta$ which best model the experimental coupling matrix, $J_{ij}$, and use those in \cref{eq: pseudo-inf potential}. 

\section{State preparation schemes} \label{app: state preparation}
In this section, we discuss how $1$-meson wave packets are engineered in both of our protocols and derive errors arising from transitions to off-resonant states. Preparing wave packets is an essential step required to demonstrate scattering in experimental platforms and is a central component of our experimental proposal. Due to the structure of 1-meson wave packets in the eigenbasis of $\hat{H}_0$ in \cref{eq:H0}, their preparation can be thought of as analogous to the preparation of W-states but with a non-uniform distribution of spin excitations. 
Both of our proposed wave-packet-preparation schemes describe an engineered resonant transition between an easy-to-prepare initial state $\ket{\psi_i}$ and a Gaussian wave packet  $\ket{\psi_g(x_0,k_0)}$ in the $\hat{H}_0$ basis as described by \cref{eq: Undressed gaussian wave packet}. 
For the blockade scheme, the initial state is the all-spin-up state. For the quantum-bus-mediated scheme, the initial state is the all-spin-up state with an excitation in the bus register. In the spin-boson implementation of the bus protocol, this translates to a boson mode initialized with one quantum. If given sufficiently long times in the experiment, these schemes can be described only by their resonant processes to the desired wave-packet state. However, in order to demonstrate scattering on current and near-term quantum simulators, time is a valuable resource and wave packets should be prepared as fast as possible while controlling errors. 

Undesired processes arise in both methods from nearby off-resonant transitions coupled to the prepared wave packet. For the blockade state-preparation scheme, the off-resonant transitions are to states with two spin flips. For the quantum-bus-mediated preparation, the off-resonant transitions are to the states with all spins up and a single excitation in the bus register. In the spin-boson implementation, additional bus registers are the set of all boson modes except the initial target mode.  The undesired states can be adiabatically eliminated if the corresponding matrix element of the off-resonant terms, $\hat{V}$ in the Hamiltonain $V_n = \bra{n} \hat{V} \ket{\psi_g(x_0,k_0)}$, connecting the nearby off-resonant state, denoted by $\ket{n}$, and the prepared wave packet is smaller than the corresponding energy difference, $\delta_n = E_n - E_g$.
In this case, the lowest-order error describing the final probability of being in the off-resonant state at time $T$ goes as $|V_n/\delta_n|^2$. The sum over all such off-resonant states will be denoted by $\varepsilon$:
\begin{equation} \label{appendix: Leading order error}
    \varepsilon \coloneqq \sum_n \abs{ \frac{V_n}{\delta_n}}^2.
\end{equation}
The subsequent sections will explicitly show how the resonant transitions are engineered and how transitions to off-resonant states scale as a function of time and wave-packet width, for both the blockade scheme and the quantum-bus-mediated scheme.

\subsection{Blockade state preparation}

Our first state-preparation protocol utilizes beyond-nearest-neighbor couplings to engineer a resonant transition to a localized wave packet. The idea behind this technique is that, by simultaneously addressing each spin using a specific drive, the beyond-nearest-neighbor interactions energetically forbid nearby excitations. If we choose the amplitude of the site-dependent driving field to be proportional to the wave-packet amplitude, then a single excitation will be distributed across the chain according to the desired wave-packet profile.

Blockade state preparation is performed using the following Hamiltonian
\begin{align} \label{eq: methods-blockade-hamiltionian}
    \hat{H}_{\rm B} = \hat{H}_0 + \hat{H}_1,
\end{align}
with $\hat{H}_0$ defined in \cref{eq:H0} and
\begin{align}
     \hat{H}_1 = -  \sum_i h^z_i \cos(\omega_i t + \phi_i) \hat \sigma^z_i .
\end{align}
The transverse field is driven at $\omega_i = E_i - E_0$ where $E_i = \bra{1,i} \hat{H}_0 \ket{1,i}$ and $E_0 = (\bra{\uparrow}^{\otimes N}_x) \hat{H}_0 \ket{\uparrow}^{\otimes N}_x$. While not necessary, it is convenient to include in $\hat H_0$ the pseudo-infinite potential $h_i^\infty$ because it introduces approximate translational invariance such that $\omega_i$ are the same for all spins. We can express the Hamiltonian in \cref{eq: methods-blockade-hamiltionian} in terms of the energy eigenstates of $\hat{H}_0$, which are just $\hat{\sigma}_x$ eigenstates. We denote these eigenstates as $\ket{\vs\,}$ where $\vs = ( s_1, s_2, \cdots, s_N)$ and $s_i \in \{ \downarrow, \uparrow \}$ (dropping the subscript $x$ on states for notational brevity), with $E_{\vs}$ denoting the energy of state $\ket{\vs\,}$. Then,
\begin{align}
    \hat{H}_0 & = \sum_{\vs} E_{\vs} \ket{\vs\,} \bra{\vs\,}, \\
    \hat{H}_1 & = - \sum_i h^z_i \cos(\omega t + \phi_i ) \sum_{\vs, \vsp}  \ket{\vs\,}\bra{\vs\,}\hat{\sigma}^z_i \ket{\vsp}\bra{\vsp}.
\end{align}
We can move into the frame rotating with $\hat{H}_0$ to obtain 
\begin{align}
    \hat{H}_{\rm R } 
     = & - \sum_i h^z_i \cos(\omega t + \phi_i ) \sum_{\vs, \vsp}    e^{i  ( E_{\vs} - E_{\vsp}) t} \nonumber\\
     & \times \ket{\vs \,}\bra{\vs\,}\hat{\sigma}^z_i \ket{\vsp}\bra{\vsp} .
\end{align}
Expanding the cosine and relabeling the energy difference as $\nu_{\vs \vsp} \coloneqq E_{\vs} - E_{\vsp}$ gives
\begin{align} \label{app: H_RWA}
    \hat{H}_{\rm R}  = &  -  \sum_i \sum_{\vs, \vsp} \frac{ h^z_i }{2} \Big( e^{i ( \nu_{\vs \vsp} + \omega ) t}e^{i \phi_i} + e^{ i ( \nu_{\vs \vsp} - \omega )t}e^{-i \phi_i}  \Big)  \nonumber\\
    & \times \ket{\vs\,}\bra{\vs\,}\hat{\sigma}^z_i \ket{\vsp}\bra{\vsp} .
\end{align}
When $\omega = \pm \nu_{\vs \vsp}$, one or the other term in the parentheses will oscillate with frequency $\pm 2\omega$ while the other will be resonant. Therefore, in the rotating frame, the state $\ket{\vsp}$ will transition to the state $\ket{\vs\,}$ only if $\ket{ \vs}$ and $\ket{\vsp}$ differ by a spin flip at site $i$ and if their energy difference $\nu_{\vs \vsp}$ is close to the driving frequency $\omega$.

For our state-preparation scheme, one needs to drive a transition from the all-spin-up state to a superposition of states with a spin down on site $i$, denoted by $\ket{1,i}$, by setting $\omega = E_1 - E_0$ (recall that we assumed $E_{i}$ are equal for all $i \geq 1$). The nearest off-resonant states are those with two spin flips at sites $i$ and $j$. For these states, the energy difference between a single-spin-flip state and a two-spin-flip state at sites $i$ and $j$ is $\omega + \delta_{ij} $ where $\delta_{ij}=4J_{ij}$ is the interaction energy between spins at sites $i$ and $j$. We can expand \cref{app: H_RWA} in terms of the states most easily accessible from the initial all-spin-up state:
\begin{align}
        \hat{H}_{\rm R}  = & - \frac{1}{2} \sum_i h^z_i \left( e^{-i \phi_i} \ket{1,i}\bra{\uparrow}^{\otimes N}_x + \rm{h.c.} \right) \nonumber \\
        & - \frac{1}{2} \sum_{i,j \neq i} h^z_j \Big(  e^{i\delta_{ij}t} e^{-i \phi_j} \hat{\sigma}^-_j \ket{1,i}\bra{1,i} + \rm{h.c.}  \Big) \nonumber \\
        &  + \cdots .
\end{align}
The first term describes the resonant transition from the all-spin-up state to the single-spin-flip states, while the second term describes off-resonant transitions from the single-spin-flip states to the two-spin-flip states. The ellipses denote all other off-resonant contributions. To prepare a Gaussian wave packet, as described in \cref{eq: Undressed gaussian wave packet}, one can set $h^z_i = \Gamma J_0 \psi^g_i(x_0, 0) $ and $\phi_i = - k_0 x_i$. Inserting this choice into the above Hamiltonian gives 
\begin{align} 
        \hat{H}_{\rm R} = & - \frac{\Gamma J_0 }{2} \left( \ket{\psi_g(x_0, k_0)}\bra{\uparrow}^{\otimes N}_x + \rm{h.c.} \right) \nonumber \\ 
        & - \frac{\Gamma J_0 }{2} \sum_{i,j \neq i} \psi_j^g(x_0, k_0) \Big(  e^{i\delta_{ij}t}  \hat{\sigma}^-_j \ket{1,i}\bra{1,i} +  \rm{h.c.} \Big) \nonumber \\
        &  + \cdots .
\end{align}
Now, the first term produces the Gaussian wave packet out of single-spin-flip states centered at $x_0$ in position space and at $k_0$ in momentum space. If driven sufficiently slowly, preparation of the Gaussian wave packet occurs at $ J_0 T = \pi/\Gamma$. It is worthwhile to note that if a given platform cannot implement the pseudo-infinite potential, then the drive frequency $\omega_i$ needs to be different at each site. To compensate for the site-dependent phase introduced by the site-dependent energy differences, an additional site-dependent phase shift is required.

In order to estimate the effect of nearby off-resonant contributions, one needs to consider the matrix element describing the transition from the wave packet to a state with two spin flips at sites $i$ and $j$, $\ket{ij}$,
\begin{align}
    \abs{V_{ij}^g} & = \bigg | \bra{ij}  \bigg( \sum_{\ell, m \neq \ell} \frac{\Gamma J_0}{2} \psi^g_{\ell}(x_0,k_0)    
    \nonumber \\
    & \hspace{1.6 cm} \hat{\sigma}_\ell^- \ket{1, m} \bra{1,m}  \bigg)\ket{\psi_g(x_0,k_0)} \bigg | \nonumber \\
    & = \bigg | \frac{\Gamma J_0}{2} \sum_{\ell,m} \psi^g_{\ell}(x_0, k_0) \psi^g_m(x_0, k_0) \left( \delta_{im} \delta_{j\ell} + \delta_{i\ell} \delta_{jm} \right) \bigg | \nonumber \\
    & = \abs{ \frac{\pi}{T} \psi_i^g(x_0, k_0) \psi_j^g(x_0, k_0) }.
\end{align}
The approximate error associated with all such off-resonant contributions according to \cref{appendix: Leading order error} is
\begin{align}
    \varepsilon_{\rm{blockade}} & \coloneqq \sum_{i,j>i} \epsilon_{ij}  = \sum_{i,j>i} \abs{\frac{V_{ij}^g}{\delta_{ij}}}^2 \\ 
    & = \sum_{i=1}^{N-1} \sum_{j=i+1}^{N} \frac{1}{T^2} \abs{  \pi \frac{ \psi_i^g(x_0, k_0) \psi_j^g(x_0, k_0) }{ \delta_{ij} }}^2.
\end{align}
There are two competing quantities that determine the behavior of this error; the interaction energy $\delta_{ij}$ and the simultaneous support of the wave packet on two sites $\abs{\psi^g_i(x_0, k_0) \psi^g_j(x_0,k_0)}$. In order to suppress this error, the wave packet should be sufficiently localized such that it does not have support across spins whose interaction energy is small. For neighboring spins, the error reduces to the condition $\Gamma \ll 4$ which limits the magnitude of the driving field $h^z$ to small amplitudes during blockade-state preparation.
For far away spins, the error is small because the wave packet does not have simultaneous support on sites $i$ and $j$, i.e., $\abs{\Gamma \psi_g^i(x_0, k_0) \psi_g^j(x_0,k_0) }^2 \ll 1$. For these reasons, the preparation of wider wave packets requires longer evolution times compared to narrower wave packets.

Preparing and scattering larger $\ell$-mesons is of significant interest, especially since larger mesons are more energetic and their scattering might lead to interesting inelastic particle production \cite{surace_scattering_2021}. In general, this would require engineering a resonant transition from some product state, most likely the all-spin-up or all-spin-down state, to these $\ell$-meson wave packets. One could imagine preparing larger mesons by expanding the blockade scheme into $\ell$ stages. First, one resonantly drives an all-spin-up state to the 1-meson manifold, then changes the driving frequency to be resonant with $\omega_2 = E_2 - E_1$ to drive the system to the 2-meson manifold and so on. During each stage, the transverse-field strength would be tailored such that the final state is a Gaussian wave packet of $\ell$-mesons. This procedure would take at least $J_0 T_{\ell} \geq \ell \tfrac{\pi}{\Gamma}$, but most likely longer since each subsequent stage should be driven slower. This is because $\ell$-mesons correspond to higher-lying states in the spectrum, which typically have smaller energy gaps to the next excited states. To respect adiabaticity, the preparation time should be slower to avoid exciting unwanted states. One could also imagine engineering a direct transition from a product state to a $2$-meson, for example, if one had access to two-body transverse terms like $\sum_i h_i^z(t) \hat{\sigma}^z_i \hat{\sigma}^z_{i+1}$. However, engineering such terms in current experiments is challenging.
\subsection{Quantum-bus-mediated state preparation}

Our second scheme applies quantum buses to the task of preparing wave packets. Starting with a state initialized in the all-spin-up state, this wave-packet-preparation scheme proceeds in two steps: (1) preparing an excitation in a chosen boson mode, (2) facilitating an exchange between the excitation in the boson mode and a wave packet in the spin register. 

\begin{figure}
    \centering
    \includegraphics[scale=0.43]{ 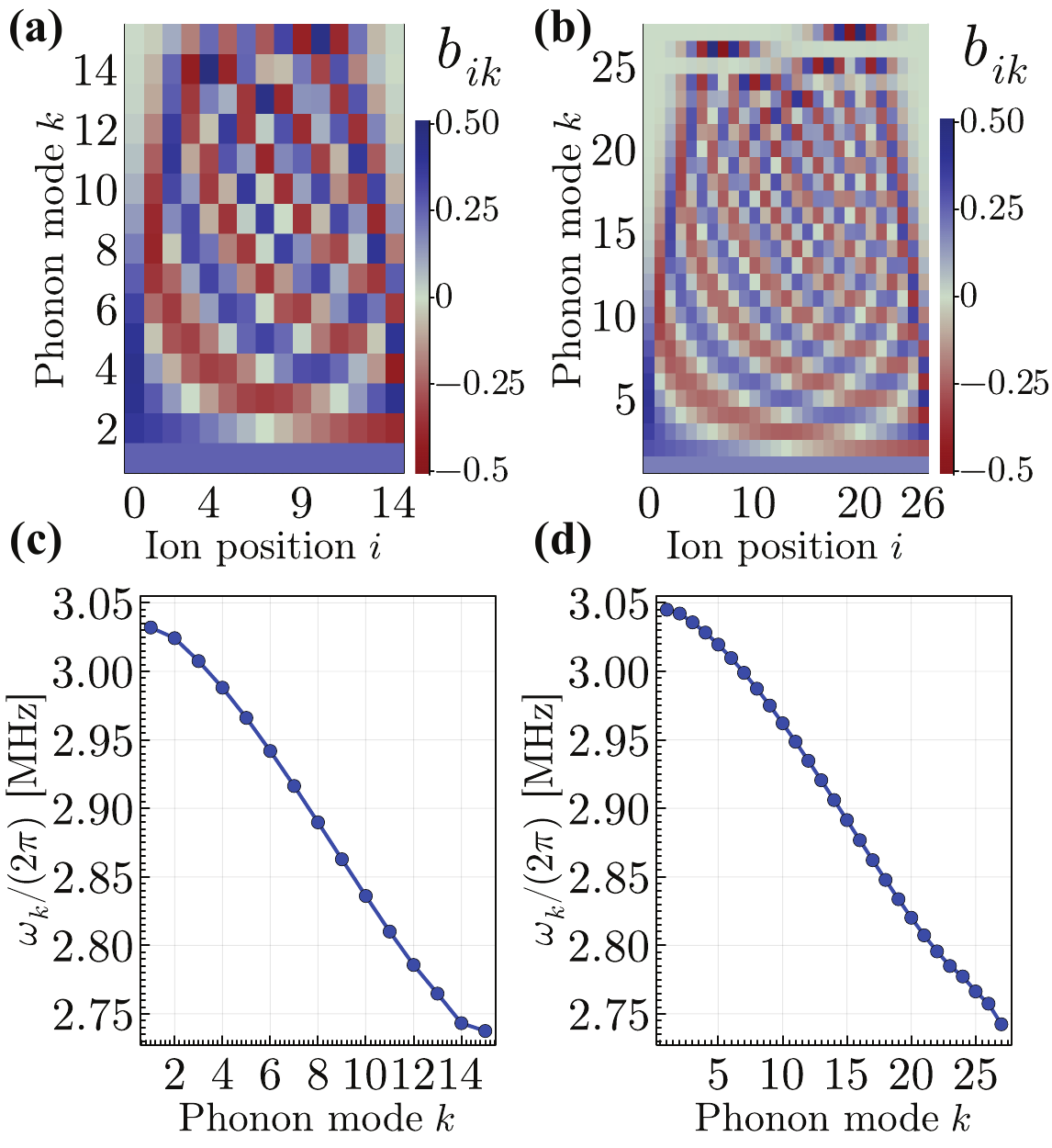}
    \caption{\textbf{Experimental details for the trapped-ion realization of quantum-bus-mediated state preparation.} Details of the experimental parameters used in benchmarking the quantum-bus-mediated state preparation realized with trapped ions as shown in \cref{fig:state-prep}(c-e). 
    Site-dependent orthonormal mode-participation matrix of the collective phonon modes for (a) $N = 15$ and (b) $N = 27$ ion chains, as well as the phonon-mode frequencies $\omega_k/(2\pi)$ for chains with (c) $N=15$ and (d) $N = 27$ ions. The simulations of this work correspond to systems with $N = 13$ and $N = 23$ spins. We take ions $1 - 13$ from the 15-ion chain and ions $1 - 23$ for the $27$-ion chain to map to these spins. The quantities plotted also contribute to the strength of the error $\varepsilon_{\rm bus}$ in \cref{eq: epsilon-bus specific}.}
    \label{fig: spin-phonon details}
\end{figure}

To prepare an excitation in a particular boson mode, a carrier transition is applied at site $j$ to flip the spin down. Then an excitation in the chosen boson mode can be created while flipping the spin back to an up state by applying the Hamiltonian in \cref{eq: spin phonon coupling}. For example, in trapped-ion systems, this can be achieved by driving predominantly the first-order blue-sideband transitions, choosing the amplitude in the anti-Jaynes-Cummings Hamiltonian in \cref{eq: spin phonon coupling} to be $A_{ik} = \eta_k b_{ik} \Omega_0/2 $ for $i = j$ and $A_{ik} = 0$ otherwise, and frequency to be resonant with the $k_t$ mode, i.e., $\nu = \omega_{k_t}$. Here, $\Omega_0$ is the Rabi frequency and $k_t$ is the mode that is aimed to be populated by one excitation. If the spin on site $j$ is down, then the above choice couples the $j^{\rm th}$ spin to the chosen target boson mode $k_t$. Evolving this system for $ T = \pi/( \Omega_0\eta_{k_t} b_{j k_t} )$ with a constant-amplitude drive $\Omega_0$ initializes a single excitation in the target boson mode and flips the $j^{\rm th}$ spin back to an up state, thus returning the spins to the all-spin-up state. Depending on the experimental implementation, each boson mode may not be uniformly coupled to all spins. For example, in trapped-ion systems, the coupling strength between a given spin and a given phonon mode is described by the orthonormal mode-participation matrix $b_{ik}$. In this case, the target mode $k_t$ should be chosen such that $b_{ik_t}$ has maximal support on spins involved in the wave packet. Due to a stronger heating for the center-of-mass and nearby long-wavelength modes \cite{katz_programmable_2023}, the target phonon mode should ideally be far away from the center-of-mass mode. Examples of experimental mode-participation matrices as well as phonon-mode frequencies for $N= 15$ and $27$ ion chains are given in \cref{fig: spin-phonon details}. 

The second (and final) stage of the process involves transferring the excitation from the target boson mode back into the spin register with the excitation distributed according to a chosen wave-packet profile. This step is described in sufficient detail in \cref{sec:proposal}, with the end result noted in \cref{eq:H-bus-mediated}. To understand the error arising from off-resonant contributions, i.e., the second term in \cref{eq:H-bus-mediated}, one needs to compute the corresponding matrix element between the Gaussian wave packet and the neighboring modes, $k \neq k_t$:
\begin{align}
    |V_k^g| = & \bigg | \bra{k}_b \bra{\uparrow}^{\otimes N}_x \bigg( \Omega_0 \sum_{i,p}  B_{ip} 
     \psi_i^g(x_0, k_0)^* 
     \nonumber \\
    & \hspace{2 cm} \sigma_i^+ a_p^\dagger \bigg)  \ket{0}_b \ket{\psi_g(x_0, k_0)} \bigg | \\
     = & \bigg | \Omega_0 \sum_{i} \abs{ \psi^g_i(x_0, k_0) }^2 B_{ik} \bigg |.
\end{align}
Here, $\ket{0}_b$ denotes the boson vacuum state while $\ket{k}_b$ denotes the state of one boson in mode $k$.
The approximate probability of undesired transitions into the non-target boson mode at $T = \pi/(2 \Omega_0) $ goes as 
\begin{align} \label{eq: epsilon-bus}
    \varepsilon_{\rm{bus}} & \coloneqq \sum_{k \neq k_t} \abs{ \frac{V^g_k}{\delta_k}}^2 \\ \label{eq: epsilon-bus specific}
     & =  \sum_{k \neq k_t} \frac{1}{ T^2} \abs{ \frac{ \pi }{2 \delta_k} \sum_{i=1}^N \abs{ \psi^g_i(x_0, k_0) }^2 B_{ik} }^2.
\end{align}
This error is 
plotted in \cref{fig:state-prep}(c) and \ref{fig:state-prep}(d). We note that this form does not entirely characterize the infidelity observed numerically. An additional error is most likely caused by a site-dependent phase shift due to the site-dependent coupling to each non-target mode. 

The scaling of the error in \cref{eq: epsilon-bus specific} as a function of system size depends on how $B_{ik}$ and $\delta_k$ change as a function of $N$. Owing to orthogonality of the mode-participation matrix $b_{ik}$, while some modes could be localized, a typical element scales with the number of ions as $b_{ik} = O(1/\sqrt{N})$ [such that $B_{ik} = b_{ik}/b_{ik_t} = O(1)$]. Assuming the trapping potential maintains a mode spectrum bandwidth that is independent of $N$, we can take $\Delta \omega_k = O(1/N)$, where $\Delta \omega_k$ is the mode spacing. The scaling of detuning from a given mode, $\delta_k$, should, therefore, be comparable to the scaling of $\Delta \omega_k$. Then, since the wave-packet state is normalized, i.e., $\sum_i \abs{\psi^g_i(x_0, k_0)}^2 = 1$, the total error scales as $\varepsilon_{\rm bus} = O\left( N^3/T^2\right)$. Therefore, increasing $T$ as $N^{3/2}$ can maintain a constant error as the chain size is increased.

There are many variations of the quantum-bus-mediated state preparation scheme that one can imagine.
First, assume we start with all spins in state down and the bosonic mode in vacuum, 
evolve for a short time under \cref{eq: spin phonon coupling}, measure the number of excitations in the bosonic mode, and find that the result is one. Then the spins would be projected onto a wave-packet state of a single delocalized up spin on top of all down spins. This approach is particularly useful if the bosonic mode is not a long-lived cavity mode, but is instead a lossy cavity mode or is replaced with a continuum of free-space modes. In fact, this process forms the basis of atomic-ensemble-based quantum repeaters~\cite{duan01a}. Second, an itinerant single photon can be mapped onto a desired wave packet using 
photon storage~\cite{hammerer10,gorshkov07a,gorshkov07b}, i.e., the reversible storage of light in atomic memory, a process that can be particularly efficient for ordered arrays~\cite{manzoni18}. Third, one can prepare spins in state $\otimes_{j=1}^N \left( \sqrt{1-p_j} \ket{\uparrow}_x + e^{i k_0 x_j} \sqrt{p_j} \ket{\downarrow}_x \right)$, engineer dispersive quantum-non-demolition coupling $\propto \hat a^\dagger \hat a \sum_i \hat \sigma^x_i$~\cite{schleier-smith10a,greve22} of the spins to the bosonic mode, and measure the energy shift of the bosonic mode. This effectively measures $ \sum_i \hat\sigma^x_i$, and if the measurement result is $N-2$, then a wave packet $\propto \sum_j e^{i k_0 x_j} \sqrt{p_j} \ket{1,j}$ of one delocalized down spin is prepared~\cite{childs02a}.

\section{Experimental details of a trapped-ion implementation} \label{app: trapped-ion experimental details}
This section outlines additional details required to realize our experimental protocol with trapped-ion quantum simulators. While the proposal of this work is suitable for many spin quantum simulators, trapped-ion platforms are particularly suited to carry out the outlined protocols in the near term. Trapped-ion simulators exhibit controllable short- and long-range Ising interactions, quantum buses realized by spin-phonon couplings, as well as favorable control across all stages of scattering. 

Trapped-ion systems can realize both power-law and exponentially decaying Ising models modelled by the function $J_{ij}={ J_0e^{-\beta(r_{ij}-1)}}/{r_{ij}^\alpha}$~\cite{Feng2023,pagano_quantum_2020}. This interaction can be realized in a native manner in trapped-ion quantum simulators with tunable coefficients $\alpha$ and $\beta$~\cite{Feng2023,de_non-equilibrium_2023}. For $\beta=0$, this model recovers the long-range Ising model with power-law coupling, where $0 \lesssim \alpha \lesssim 3$ depends on the detuning of the applied optical fields from the first phonon side band of the transverse center-of-mass mode~\cite{RevModPhys.93.025001,liu_confined_2019}. For $\alpha = 0$ and $\beta > 0$, this model recovers exponentially decaying couplings that can be realized, e.g., by coupling primarily to the low-frequency transverse zig-zag modes with $\beta > 0$~\cite{nevado2016hidden,Feng2023,schuckert2023observation,katz2024observing,kim2009entanglement,lee2016floquet}.  

In trapped-ion platforms, the quantum-bus-mediated preparation realized with spin-phonon coupling offers a more favorable preparation time compared with the blockade state preparation considering the coherence time of the experiment. The native time-dependent Hamiltonian describing the spin-phonon coupling in trapped-ion platforms is given by the anti-Jaynes-Cummings Hamiltonian in \cref{eq: spin phonon coupling}, which is realized by driving the first blue side-band transitions of a single set of transverse or longitudinal motional modes. 
The site- and mode-dependent amplitudes are $A_{ik} = \eta_k b_{ik} \Omega_i/2 $, where $\eta_k$ is the Lamb-Dicke parameter of mode $k$ and $b_{ik}$ is the site-dependent orthonormal mode-participation matrix. The parameters $\eta_k$, $b_{ik}$, and phonon-mode frequencies $\omega_k$ depend on the ion trapping potential~\cite{wineland1998experimental,RevModPhys.93.025001}. To program the evolution to the desired wave packet from a chosen target mode, the site-dependent Rabi frequency is chosen to be
\begin{equation}
\Omega_i = 2 \Omega_0  \frac{\psi_i^g(x_0,k_0)}{\eta_k b_{ik_t}}.
\end{equation}
In the numerical simulations shown in \cref{fig:state-prep}(c-e), we have set $\Omega_0 = \pi/(2T)$ MHz, where $T$ is the preparation time in microseconds. Furthermore, the parameters $\eta_k \approx 0.08$, $\omega_k$, and $b_{ik}$ are used assuming $^{171}\rm{Yb}^+$ ions and a combination of quadratic and quartic electrostatic trapping potentials in the axial direction that renders the spacing between the center ions nearly equidistant~\cite{katz_demonstration_2023}. The $\omega_k$ and $b_{ik}$ values used in the $15$-ion and $27$-ion systems (corresponding to $N = 13$ and $N=25$ simulations) are shown in \cref{fig: spin-phonon details}. In trapped-ion experiments, the tunable parameter $\Omega_i$ is typically of the order $\lesssim 1~$MHz. Preparation times shown in \cref{fig:state-prep}(c-e) for $N = 13,~23$ correspond to $\Omega_i$ values in the range $0.1-10~$MHz. Here, we have estimated $\Omega_i \approx \sqrt{N} \pi / (\eta T)$, where we have used $b_{ik} = O(1/\sqrt{N})$. Therefore, a range of required Rabi frequencies are feasible in current experiments.

\bibliography{Meson-Scattering}

\end{document}